\documentclass[aps,showpacs,superscriptaddress,groupedaddress,footnote,nofootinbib]{revtex4}  
\usepackage{graphicx}  
\usepackage{dcolumn}   
\usepackage{bm}
\usepackage{amssymb}   
\usepackage{amsmath}
\usepackage{xcolor}
\usepackage{rotating}
\usepackage{amsthm}
\usepackage{subfig}
\usepackage{subfloat}
\usepackage{actuarialsymbol}
\usepackage{braket}
\usepackage{mathtools}
\usepackage{bbold}
\usepackage{amssymb}
\usepackage{enumitem}
\usepackage{comment}
\newcommand{\be}{\begin{equation}}\newcommand{\ee}{\end{equation}}
\newcommand{\bea}{\begin{eqnarray}}\newcommand{\eea}{\end{eqnarray}}
\newcommand{\brr}{\begin{array}}\newcommand{\err}{\end{array}}
\newcommand{\bit}{\begin{itemize}}\newcommand{\eit}{\end{itemize}}
\newcommand{\ben}{\begin{enumerate}}\newcommand{\een}{\end{enumerate}}

\newcommand{\ba}{\begin{array}}
\newcommand{\ea}{\end{array}}
\newcommand{\ide}{1\hspace{-1mm}{\rm I}}

\definecolor{darkred}{rgb}{.8,0,0}

\definecolor{darkblue}{rgb}{0,0,.7}

\def\al{\alpha}

\def\te{\theta}

\def\1{{_{1}}}\def\2{{_{2}}}

\def\noHe0{:\;\!\!\;\!\!:H_e(0):\;\!\!\;\!\!:}
\def\noHm0{:\;\!\!\;\!\!:H_\mu(0):\;\!\!\;\!\!:}
\def\al{\alpha}

\def\te{\theta}

\def\1{{_{1}}}\def\2{{_{2}}}
\DeclareMathOperator{\Tr}{Tr}

\begin{document}
\title{Complete complementarity relations in tree level QED processes}

\author{Massimo Blasone, Silvio De Siena\footnote{Universit\`a degli Studi di Salerno (Ret. Prof.),  email: silvio.desiena@gmail.com }, Gaetano Lambiase, Cristina Matrella and Bruno Micciola}
\email{blasone@sa.infn.it}\email{lambiase@sa.infn.it}
\email{cmatrella@unisa.it}\email{bmicciola@unisa.it}
\affiliation{Dipartimento di Fisica, Universit\`a di Salerno, Via Giovanni Paolo II, 132 I-84084 Fisciano (SA), Italy}
\affiliation{INFN, Sezione di Napoli, Gruppo collegato di Salerno, Italy}

\date{\today}

\begin{abstract}

We exploit the complete complementarity relations (CCR) to fully characterize various aspects 
of quantumness in QED scattering processes at tree level. As a paradigmatic example, we consider 
Bhabha scattering  in two different configurations: 
in the first case, the initial state is factorized in the spin 
and we study the generation of entanglement  due to the scattering. 
Then we consider the most general case in which the initial state can be entangled: 
we find that the scattering generates and distributes quantum information 
in a non-trivial way among the spin degrees of freedom of the particles, 
with CCR relations being preserved. An important outcome of our analysis 
is that maximal entanglement is conserved in the scattering process 
involving only fermions as input and output states, with a more complex situation if photons are present.  
\end{abstract}

\vskip -1.0 truecm

\maketitle

\section{Introduction}

In the last 25 years the development of quantum information theory
has highlighted the pivotal role of the concept of entanglement,  describing correlations of exclusive quantum nature,  
as the watershed with respect to  classical phenomena where such kind of correlations cannot be generated at all.
Entanglement is the object of very active research, due to its conceptual relevance 
in basic science and for being a resource for quantum technologies, 
such as quantum teleportation \cite{Ren}, quantum cryptography \cite{Yin}, quantum computation \cite{Jozsa}
and quantum metrology \cite{Demkowicz}. Parallel to the development of quantum information, 
entanglement has proved very useful in other scientific areas, as the study and characterization
of quantum phase transitions \cite{Osterloh}
and, in quantum field theory (QFT), scaling laws \cite{Eisert:2008ur} and conformal field theories \cite{Calabrese:2004eu}.
The idea that entanglement can be exploited as a probe to investigate the deepest quantum nature and  hidden mechanisms of
fundamental interactions is gaining ground more and more. 
In mesonic (in particular, kaonic) systems, violations of Bell inequalities 
and presence of entanglement have been connected with discrete symmetries 
and special relativity probes \cite{Benatti}-\cite{Bernabeu:2015aga}, and in QCD entanglement suppression has been related
with approximated spin-flavor symmetries \cite{Beane:2018oxh}. 
Entanglement in top-antitop quark events produced at LHC has been recently observed \cite{Afik:2022dgh,ATLAS:2023fsd}.
Quantum correlations, including entanglement, have also been extensively investigated 
in the context of neutrino oscillations \cite{Blasone:2010ta}-\cite{Bittencourt:2023qik}. 
Another line of investigation concerns the  role of entanglement in particle processes  
as a probe for physics beyond standard model  \cite{Duch}-\cite{Manoukian:2004xt}. 
Finally, a vast literature has been devoted to investigate hidden contents of non-classicality in Gravity \cite{Ryu}-\cite{Balasubramanian:2011wt}.

A great attention has been recently devoted to the study of entanglement in QFT scattering and decay processes 
as in Refs. \cite{Peschanski:2016hgk}-\cite{Blasone:2024dud}. 
In particular, a systematic investigation of the entanglement generation  
and distribution in fundamental processes has been recently developed 
in Refs. \cite{Cervera-Lierta:2017tdt, CerveraLierta:2019ejf} and \cite{Blasone:2024dud}.
In Refs. \cite{Cervera-Lierta:2017tdt, CerveraLierta:2019ejf},  
entanglement has been studied at tree level in QED and  in electroweak processes, 
with the aim of showing that the constraint on the coupling structure 
of QED and weak interactions introduced by requiring maximal entanglement 
between helicity degrees of freedom,
can lead to extract significant physical aspects of these fundamental theories. A systematic analysis 
of helicity entanglement generation in QED scattering processes has been carried out 
in Ref. \cite{Fedida:2022izl}-\cite{Fedida:2024dwc} and, in particular, in Ref. \cite{Fedida:2022izl} it has been studied for all energies and for arbitrary initial mixtures of helicity states. 
The analysis was performed at tree-level and one-loop contributions were estimated to be small.

In this paper we perform a complete study of entanglement in helicity for QED scattering processes at tree level. 
We do this by exploiting the so-called complete complementarity relations (CCR) 
which have already shown to be very useful in characterizing neutrino oscillations 
\cite{Bittencourt1,Bittencourt:2023qik}. The CCR incorporate into a conservation constraint all the complementary aspects 
which can be associated to a bipartite quantum system 
(in a pure or mixed state): predictability,
local coherence (visibility) and nonlocal correlation (nonlocal coherence). 
The CCR establish that the sum of the squares of these three quantities remains constant, 
and provide a tool to study the generation and transformation of such three quantities in consequence of a physical interaction.
Equipped with this tool, we show that a QED scattering process modifies and distributes 
quantum information in a non-trivial way  among the spin degrees of freedom of the particles, 
with the CCR fulfilled both for initial and final states. 

We concentrate our analysis on the Bhabha scattering process as paradigmatic case, 
and we study other QED processes for the particularly interesting instance of initial maximally entangled states. 
For the simplest case of an initial factorized state, 
as e.g. $\ket{RL}$ where $R,L$ indicate helicity eigenstates, 
generation of entanglement has been already discussed in Ref.\cite{Cervera-Lierta:2017tdt}: 
however, the use of CCR allows us to provide a complete characterization of the final state 
by revealing that entanglement generation is indeed related 
to the interplay between the three terms of CCR, depending on the scattering parameters. 

We proceed in our analysis by considering more complex input states, 
which include an initial amount both of predictability and of local coherence in the two single-particle systems. 
Finally, we consider the most general initial state, in which the incoming particles can be entangled. 
In this case we can summarize the principal results as follows. 
If the initial state is maximally entangled, the QED processes 
involving only fermions as input and output states preserve maximal entanglement,
while a more complex situation
occurs if also photons are present. 
In the case of non-maximally entangled initial states we find  behaviors 
in which, for a particular class of states, the entanglement is greater than the initial one, while for others it can be partially, or even totally, suppressed for any scattering angle (we define such regimes as \textit{entanglophilus} 
and \textit{entanglophobus}, respectively). For other configurations mixed behavior is detected.  
Finally, we have estimated  entanglement as a \emph{quantum resource} generated in the scattering process, 
by considering the average CCR terms per particle in a given domain of the scattering angle. 

The paper is organized as follows. 
In section \ref{CCR} we briefly review the complete complementarity relations. 
In section \ref{CCRBhabha} we study Bhabha scattering with the CCR relations for the different classes of initial states. 
In section \ref{MES} we analyze the outputs of initial maximally entangled states
for various QED processes. 
In section \ref{ER} we show the different classes in which can be partitioned the non-maximally entangled states.
In section \ref{entres} we study the entanglement as a resource in various regions of the scattering angle,
and in section \ref{Disc} we provide some interpretation.
Finally, in section \ref{conclusions}, we discuss the results along with conclusions and outlook.


\section{Complete complementarity relations}
\label{CCR}

In 1927 Niels Bohr conceived his complementarity principle \cite{Bohr}, 
which states that a quantum system may possess properties which are equally real but mutually exclusive, 
in the sense that more information one has about one aspect of them, the less information can be obtained about the other. 
The concept of complementarity is often associated with wave-particle duality, 
the complementarity aspect between propagation and detection of a ``quanton" \cite{Levy}.

The first quantitative version of Bohr's principle, 
originally formulated in the context of the two-slit experiment, 
is due to Wootters and Zurek \cite{WoottersZurek}, who studied the effect of introducing 
a path-detecting device in a two-slit interference experiment, as originally proposed by Einstein. They showed that, by obtaining partial information about the quanton's path, 
the interference pattern was only partially destroyed. 
The extension of this work by Englert \cite{Englert} led to the duality relation:
 \begin{equation}
 	D^{2}+V^{2}\le 1,
 	\label{0.1}
 \end{equation}
where $D$ represents the \textit{path-distinguishability}, 
which measures the particle content, and $V$ is the \textit{visibility} of the interference. 
This led to the following relation:
 \begin{equation}
 	P^{2}+V^{2}\le 1,
 	\label{1}
 \end{equation}
 where $P$ is called \textit{predictability}, representing a measure of the path information 
 which still quantifies the particle aspect, and $V$ is the visibility before defined.

Thus, the concept of complementarity associated with wave-particle duality 
is related to mutually exclusive properties of single-partite quantum systems.
Eqs.(\ref{0.1}) and (\ref{1}) are only saturated for single-partite 
 pure states, 
while for mixed states the strict inequality holds. 
In fact, Jacob and Bergou \cite{JB1,JB2,JB3} have shown that for bipartite states a third entry, 
represented by a measure of correlation $C$ between subsystems, has to be considered. 
This is because in general, even if the global state is pure, the states of subsystems can be mixed, 
implying a lack of information about the single subsystem: the missing information 
must be recovered in the correlations between subsystems. 
In this way it is possible to obtain a \textit{complete complementarity relation} 
(CCR) for pure states, given by:
 \begin{equation}
 P_{k}^{2}+V_{k}^{2}+C^{2}=1,
  \label{2}
 \end{equation}
where $P_{k}$ and $V_{k}$ are the predictability and the visibility referred 
to the single subsystem $k$, with $k=1,2$ and $C$ is a measure of non-local correlation. 
In Ref.\cite{JB3} $C$ represents the concurrence, a measure of entanglement. 
\\

Let us consider a general two-qubit pure quantum state:
\begin{equation}
\ket{\psi}=a\ket{00}+b\ket{01}+c\ket{10}+d\ket{11},
\label{3}
\end{equation}
with $|a|^{2}+|b|^{2}+|c|^{2}+|d|^{2}=1$.
%
%
The predictability $P_{k}$ and visibility $V_k$ are defined, respectively, as: 
\bea\label{predictG}
P_{k}&=&|\langle \psi|\sigma_{z,k}|\psi\rangle |, \hspace{1cm} \sigma_{z,k}=\begin{pmatrix}
1&0\\
0&-1
\end{pmatrix},
\\
\label{VisibG}
V_{k}&=&2|\langle \psi|\sigma_{k}^{+}|\psi\rangle |, \hspace{1cm}
\sigma_{k}^{+}=\begin{pmatrix}
0&1\\
0&0
\end{pmatrix},
\eea

Finally, the concurrence is defined as $C=|\langle\psi|\tilde{\psi}\rangle|$, 
where $\tilde{\psi} = (\sigma_{y}\otimes\sigma_{y})\ket{\psi^{*}}$, $\sigma_{y}$ 
is the Pauli matrix and $\ket{\psi^{*}}$ is the complex conjugate of $\ket{\psi}$.
For the state in Eq.(\ref{3}) these quantities are given by:
\bea
&& P_{1}=|(|c|^{2}+|d|^{2})-(|a|^{2}+|b|^{2})|\,, \qquad
P_{2}=|(|b|^{2}+|d|^{2})-(|a|^{2}+|c|^{2})|.
\label{4}
\\
&& V_{1}=2|ac^{*}+bd^{*}|\,, \qquad 
V_{2}=2|ab^{*}+cd^{*}|.
\label{5}
\\
&& C(\psi)=2|ad-bc|.
\label{6}
\eea

It is simple to verify from Eqs.(\ref{4})-(\ref{6}) that the complete complementarity relation Eq.(\ref{2}) 
is satisfied for a general pure bipartite state $\ket{\psi}$.

Thus, we deal with a \textit{triality relation} formed by two quantities 
generating \textit{local} single-partite realities which can be related 
to wave-particle duality, and a correlation measure which generates an exclusive bipartite \textit{non-local} reality. 
CCR as in Eq.(\ref{2}) are valid only if the whole system is in a pure state. When a mixed state is considered the CCR becomes:
\begin{equation}
P_{k}^{2}+V_{k}^{2}+C^{2}\le1.
  \label{0.2}
 \end{equation}
This can be explained by considering that the correlation measure and the visibility 
for pure states represent upper bounds for the corresponding quantities for mixed states.

Recently, in Refs. \cite{Basso:2020,Basso:2021whv,Basso:2021cuh}  CCR have been reformulated and extended 
in a form which is particularly suitable for our purposes.
We then briefly review, for the case of a bipartite pure state, the main aspects of this approach which efficiently expresses CCR in terms of the density matrix of the system. 

Let us consider a bipartite pure state in the Hilbert space $\mathcal{H}_{AB}=\mathcal{H}_{A}\otimes\mathcal{H}_{B}$ 
of dimension $d=d_{A}d_{B}$, where $d_{A}$ and $d_{B}$ are the dimensions 
of $\mathcal{H}_{A}$ and $\mathcal{H}_{B}$, respectively. 
The density matrix associated to this state can be written as:
\begin{equation}
\rho_{AB}=\sum_{i,k=0}^{d_{A}-1}\sum_{j,l=0}^{d_{B}-1}\rho_{ij,kl}\ket{i,j}_{AB}\bra{k,l}.
    \label{0.3}
    \end{equation}
By tracing over B we obtain the reduced density matrix for subsystem A:
\begin{equation}
\rho_{A}=\sum_{i,k=0}^{d_{A}-1}\rho_{ik}\ket{i}_{A}\bra{k}.
    \label{0.4}
\end{equation}
It is important to point out that, on the basis of measures we choose 
for predictability and visibility and consequently for the correlation term, 
we can define different forms of CCR. As an example, here we show two possible forms:
\begin{equation}
    P_{hs}(\rho_{A})+C_{hs}(\rho_{A})+C_{hs}^{nl}(\rho_{A|B})=\frac{d_{A}-1}{d_{A}},
    \label{0.5}
\end{equation}
\begin{equation}
    P_{vn}(\rho_{A})+C_{re}(\rho_{A})+S_{vn}(\rho_{A})=\log_{2}d_{A}.
    \label{0.6}
\end{equation}
In Eq.(\ref{0.5}) predictability, visibility and nonlocal coherence are expressed in terms of the Hilbert-Schmidt measures, given by 
\begin{eqnarray}
    P_{hs}(\rho_{A})&=&\sum_{i=0}^{d_{A}-1}(\rho_{ii})^2-\frac{1}{d_{A}},
    \label{predbm}
    \\
    C_{hs}(\rho_{A})&=&\sum_{i\ne k}^{d_{A}-1}|\rho_{ik}|^2,
    \label{visbm}
    \\
    C_{hs}^{nl}(\rho_{A|B})&=&\sum_{i\ne k, j \ne l}|\rho_{ij,kl}|^{2}-2\sum_{i\ne k, j<l}Re(\rho_{ij,kj}\rho^{*}_{il,kl}).
    \label{entbm}
\end{eqnarray}

In Eq.(\ref{0.6}), the predictability is expressed in terms of the von Neumann entropy 
as $P_{vn}(\rho_{A})=\log_{2}d_{A}-S_{vn}(\rho_{A,diag})$ 
and $C_{re}(\rho_{A})=S_{vn}(\rho_{A,diag})-S_{vn}(\rho_{A})$ 
is the relative entropy of coherence. $S_{vn}(\rho_{A})$ is the von Neumann entropy of entanglement.

Referring to the entropic form of CCR given in Eq.(\ref{0.6}), 
for completeness we briefly show also the CCR relation for the case of a global mixed state. 
In this case $S_{vn}(\rho_{A})$ cannot be considered as a measure of entanglement between A and B, 
but it just represents a measure of uncertainty of A. Then, 
the correlation between A and B is quantified by the sum of two terms, 
given respectively by the mutual information $I_{A:B}(\rho_{AB})$, 
a measure of the total correlation between A and B, and by the conditional entropy $S_{A|B}(\rho_{AB})$, 
which quantifies the ignorance about the whole system one has by looking only at subystem A. 
Thus, for a global bipartite mixed state, the entropic form of CCR becomes:
\begin{equation}
     P_{vn}(\rho_{A})+C_{re}(\rho_{A})+S_{A|B}(\rho_{AB})+I_{A:B}(\rho_{AB})=\log_{2}d_{A}.
     \label{0.7}
\end{equation}

Finally, for the general bipartite state Eq.(\ref{3}),
we add some more information on the form of
the concurrence, on its relation with entropy measures 
(linear $S_{lin}$ and von Neumann $S_{vn}$ entropies), and on
the maximum entanglement condition.
If we write the coefficients in Eq.(\ref{3}) in the following way:
\be
a \equiv |a|\,, \,\,\,\, b = |b| \, e^{i \xi}\,, \,\,\,\, c = |c| \, e^{i \eta}\,, \,\,\,\, d = |d| \, e^{i \tau},
\ee
we obtain, for the squared concurrence and  the quantum entropies the following expressions:
\be
C^2 = 4 \, \{|a|^2 \, |d|^2 + |b|^2 \, |c|^2
- 2 \; |a| \, |b| \, |c| \, |d| \; \cos{[\xi - (\eta + \tau)]}\}.
\ee
\begin{equation}
    S_{lin} = 2\bigl(1-Tr(\rho^2)\bigr)= C^2,
    \label{linent}
\end{equation}
\be
S_{vn} = -\rho \log_2 \rho = - \frac{1}{2} (1 + \sqrt{1 - C^2}) \,  \log_2 \, \left[\frac{1}{2} (1 + \sqrt{1 - C^2})\right]
- \frac{1}{2} (1 - \sqrt{1 - C^2}) \, \log_2 \, \left[\frac{1}{2} (1 - \sqrt{1 - C^2})\right].
\ee
\noindent We see that the more general condition for maximum concurrence is
\be
\label{condizCmax}
|a|^2 = |d|^2, \hspace{0,3cm} |b|^2 = |c|^2, \qquad \, \xi - (\eta + \tau) = (2n+1) \pi, \; n \in \mathcal{Z},
\ee
and that the maximum concurrence corresponds both to maximum linear entropy and to the maximum von Neumann entropy.


\section{CCR in Bhabha scattering process}
\label{CCRBhabha}

We start by studying in detail the three quantities above introduced (predictability, 
local coherence and entanglement) for the specific case of Bhabha scattering 
at tree level for which entanglement generation was studied in Ref.\cite{Cervera-Lierta:2017tdt}. 

In this Section we briefly recall the formalism used in Ref.\cite{Araujo} and Ref.\cite{Blasone:2024dud}. 
From now on, we perform our calculations in the COM reference frame for particles $A$ and $B$.

The internal product of fermion states is defined as
\begin{equation}
\label{intprod}
    \braket{k,a|p,b} = 2E_{\bf k}(2\pi)^3 \delta^{(3)}({\bf k}-{\bf p})\delta_{a,b},
\end{equation}
where $k$ and $p$ are the 4-momenta and $a$ and $b$ are the spin indices.

For a generic initial state the final states can be expressed in terms 
of the scattering amplitudes $\mathcal{M}(p_1,a,p_2,b\,;\, p_3,r,p_4,s)$. 
For example, considering the simple initial state:
\begin{equation}
\label{instate}
    \ket{i} = \ket{p_{1},a}_A \otimes \ket{p_{2},b}_B,
\end{equation}
the general final state results
\begin{eqnarray}
\label{finstate} \nonumber
    \ket{f} &=& \ket{i} + 
		i \sum_{r,s}\int \frac{d^3 {\bf p_3} d^3{\bf  p_4}}{(2\pi)^6 2E_{\bf p_3}2E_{\bf p_4}}\delta^{(4)}(p_1+p_2-p_3-p_4)
		\Big[\mathcal{M}(a,b;r,s)\ket{p_3, r}_A \otimes \ket{p_4, s}_B\Big],
\end{eqnarray}
where $\mathcal{M}(a,b;r,s)$ represents the scattering amplitudes 
in which for simplicity we omit the initial and final momenta.

The partial trace operation is given by
\begin{equation}
\label{trace}
    Tr_X[{\rho}] = \sum_{\sigma}\int \frac{d^3\bf k}{(2\pi)^32E_{\bf k}} 
		\bigr(\ide_r \otimes \prescript{}{X}{\bra{k,\sigma}}\bigl)\rho\bigr(\ide_r \otimes \ket{k, \sigma}_X\bigl),
\end{equation}
where $\ide_r$ denotes the identity operation in the remaining subspaces, 
$k$ and $\sigma$ are the 4-momentum and spin indices as before 
and $X$ is the generic space with respect to which we calculate the trace.
So, we can describe the final states in terms of the density matrix:
\begin{equation}
\label{rhofin}
    \rho_{AB}^{f} = \frac{1}{\mathcal{N}}\, \ket{f}\bra{f},
\end{equation}
where $\mathcal{N}$ is the normalization constant.
Using Eq.\eqref{trace} and applying the following relations
\begin{equation}
\label{Tfactor}
    2\pi\delta^{(0)}(E_i-E_f) = \int_{-T/2}^{T/2} e^{i(E_i-E_f)t} dt,
\end{equation}

\begin{equation}
\label{Vfactor}
        (2\pi)^3\delta^{(3)}({\bf k}-{\bf p}) =  V\delta_{\bf{k},\bf{p}} \hspace{0.1cm},
\end{equation}
which imply that $(2\pi)\delta^{(0)}(0)=T$ and $(2\pi)^3\delta^{(3)}(0)=V$, 
we can easily compute the normalization constant $\mathcal{N}$:
\begin{equation}
\label{norm}
    \mathcal{N} = Tr_A\Bigl[Tr_B\Bigl[\,\ket{f}\bra{f} \Bigr]\smallskip\Bigr].
\end{equation}

In this work we study the change of the CCR's terms, relative to the spin degrees of freedom, between states before and after the scattering process at fixed momenta. as in Refs.\cite{Cervera-Lierta:2017tdt}-\cite{Fedida:2022izl}, where it is assumed that an arbitrarily sharp filtering of the outgoing particles in momentum space is performed, without resolving 
their internal (helicity or polarisation) degrees of freedom. As remarked in Ref.\cite{Fedida:2022izl}, although being an idealisation, the filtering corresponds to a selection of the output momenta, and the entanglement between the helicity degrees of freedom at that given momenta is still a fundamental property of the output state of the quantum field upon scattering. So, from now on, we will consider these states in terms of helicity states.

\subsection{Factorized and polarized incoming states}

We start with the simplest instance of an initial factorized state of two  polarized particles:
\begin{equation}
\label{instateI} 
    \ket{i}_I = \ket{R}_A\ket{L}_B.
\end{equation}

After the scattering, if we limit our attention to a selection of results at each fixed angle $\theta\neq 0,2\pi$, 
up to a normalization factor, we can express 
the \emph{final reference state} as \footnote{The normalization can be fixed after the operation of momentum filtering 
(i.e. selection of the measurements relative to a specific scattering angle $\theta$) 
that is formally described by applying a POVM operator as discussed in Ref.\cite{Fedida:2022izl}.}
\begin{equation}
\label{finstateI} 
    \ket{f}_{I} = \sum_{r,s}\mathcal{M}(RL;rs)\ket{r}_A\ket{s}_B,
\end{equation}
where $\mathcal{M}(RL;rs)$ are scattering amplitudes whose explicit form is reported in the Appendix.

\begin{figure}[t]
 \subfloat[][\emph{}]{\includegraphics[width =5.55 cm]{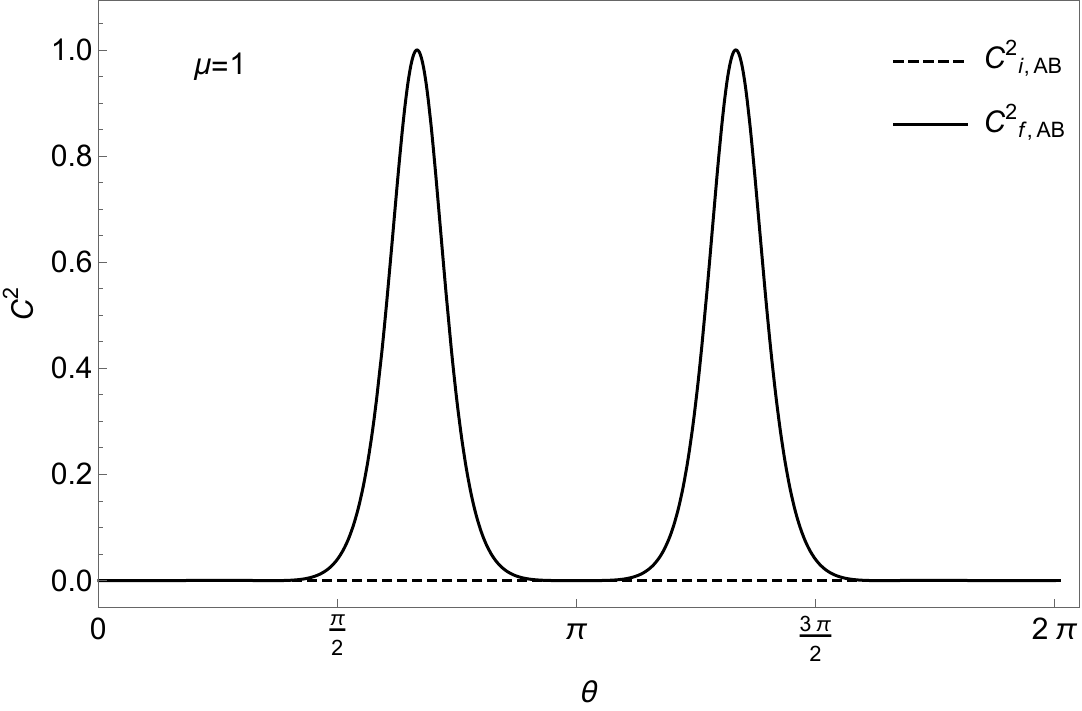}
\label{fig1a}}\quad
 \subfloat[][\emph{}]{\includegraphics[width =5.55 cm]{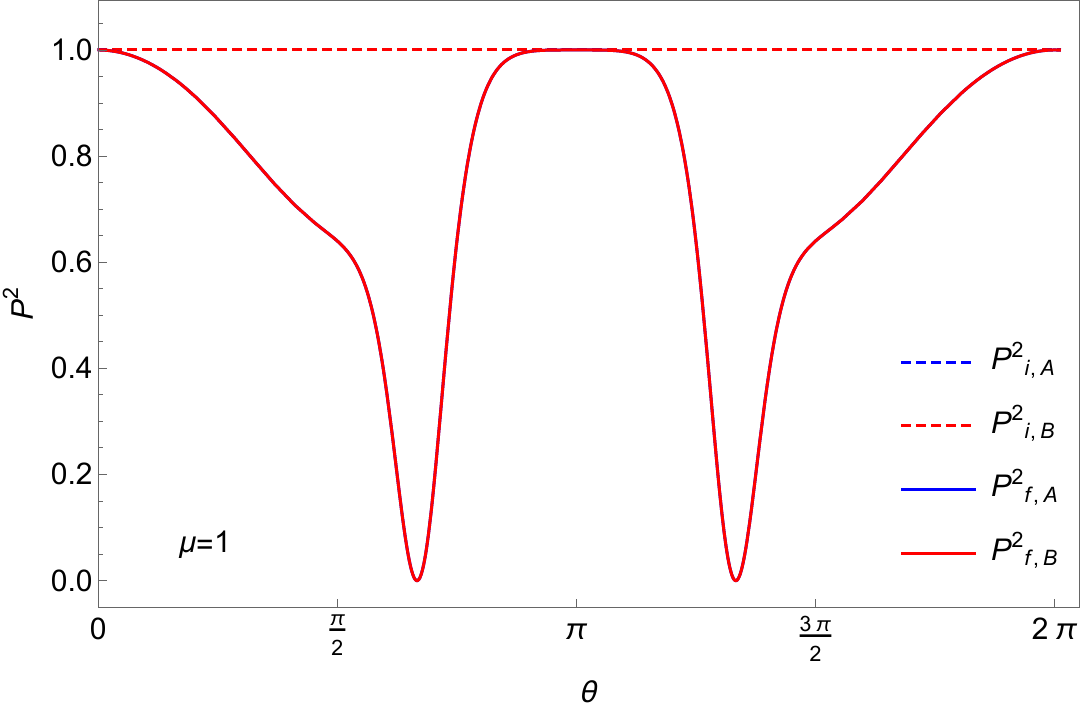}
\label{fig1b}}\quad
 \subfloat[][\emph{}]{\includegraphics[width =5.55 cm]{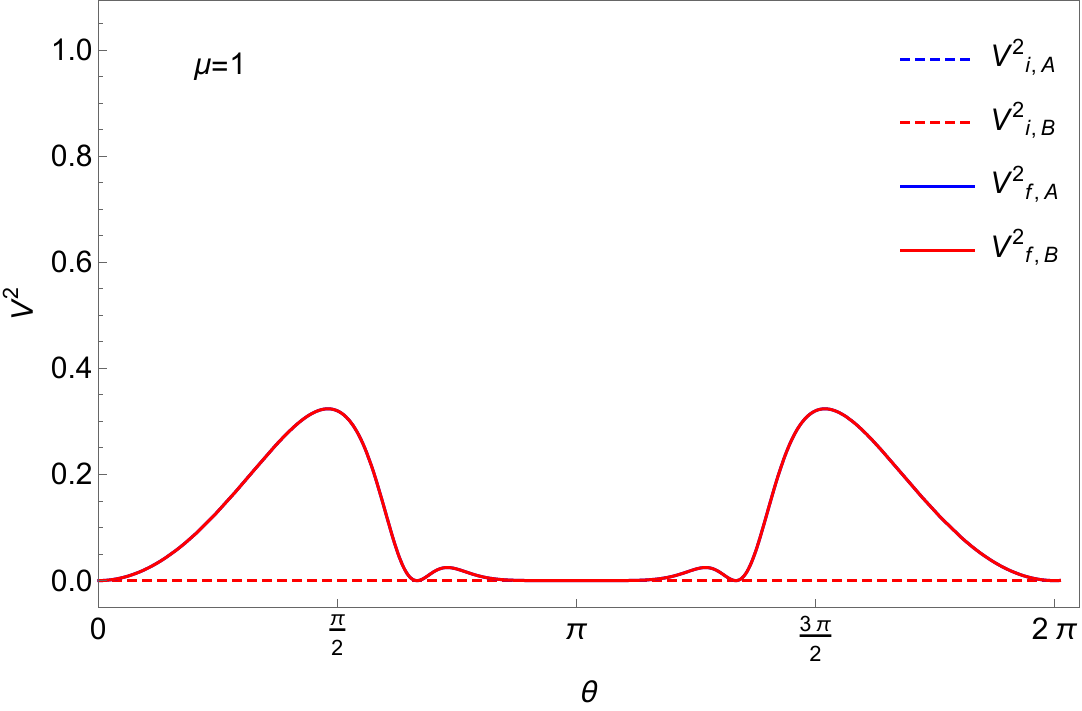}
\label{fig1c}}\quad
 \subfloat[][\emph{}]{\includegraphics[width =5.55 cm]{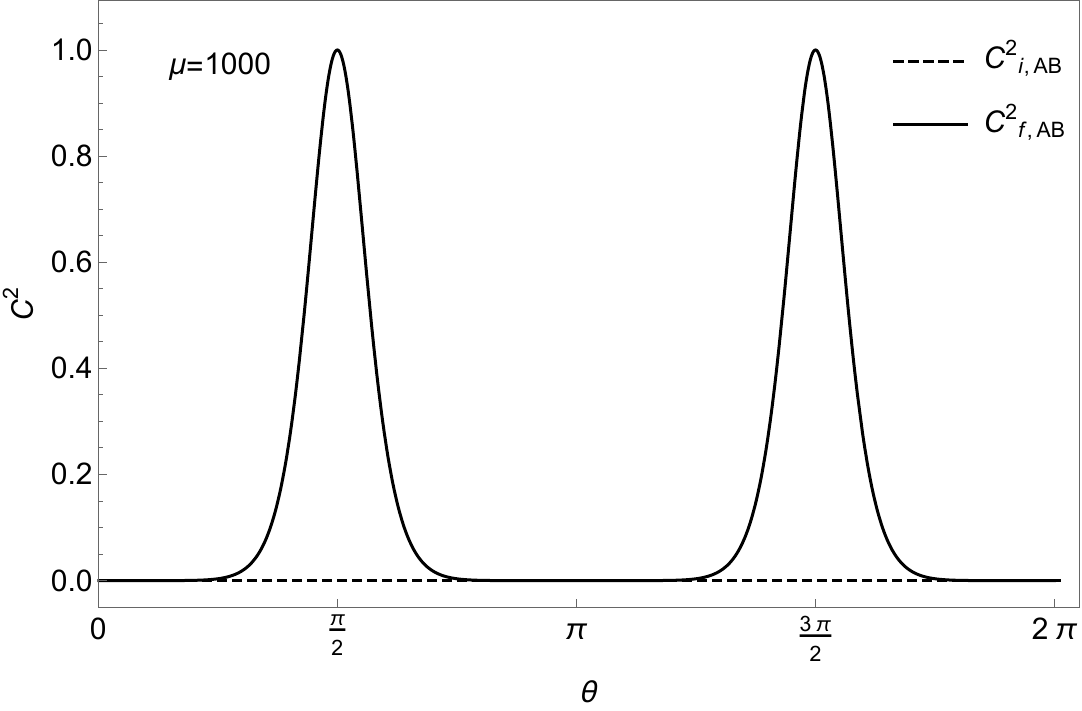}
\label{fig1d}}\quad
 \subfloat[][\emph{}]{\includegraphics[width =5.55 cm]{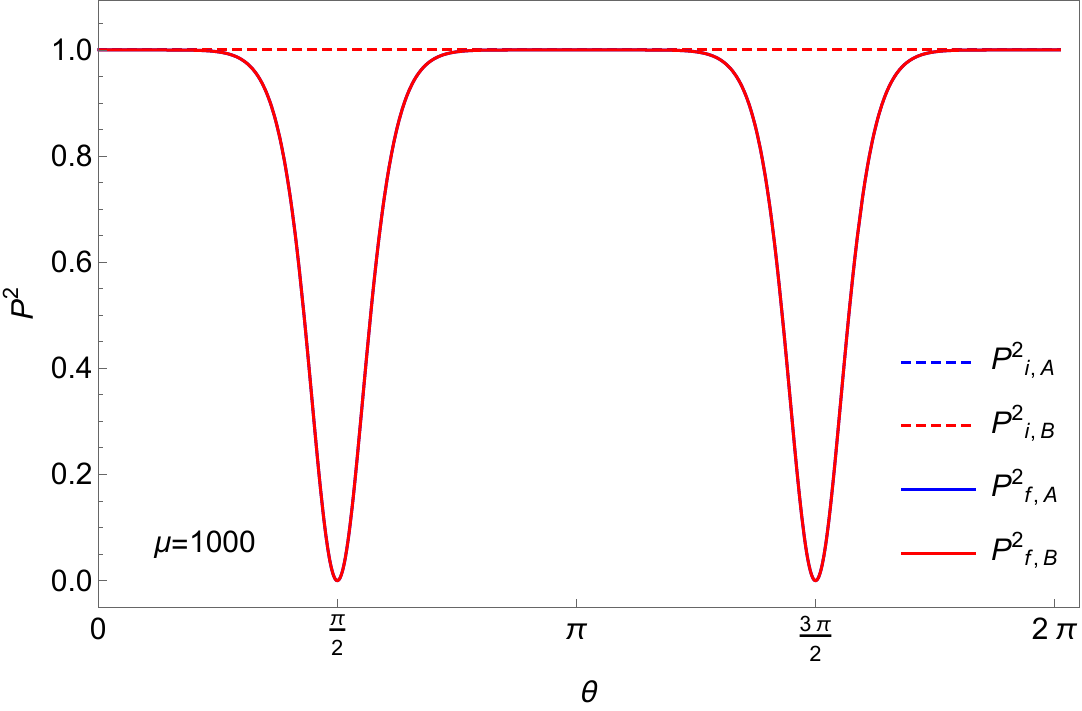}
\label{fig1e}}\quad
 \subfloat[][\emph{}]{\includegraphics[width =5.55 cm]{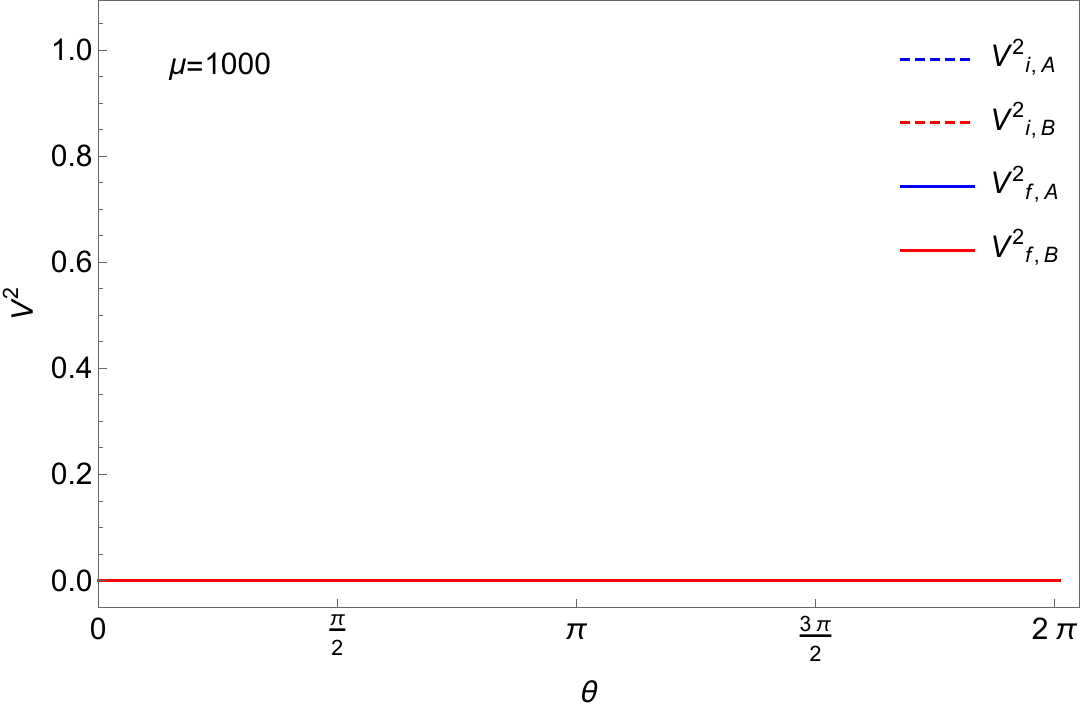}
\label{fig1f}}\quad
\caption{Plot of the CCR terms $C^2$, $P^2$ and $V^2$, for initial and final states of Case I.
Panels (a),(b),(c) refer to the value $\mu=1$ (non-relativistic regime). 
Panels (d),(e),(f) refer to the value $\mu=1000$ (relativistic regime). }  
\label{figone}
\end{figure}

The initial state $\ket{i}_{I}$ is characterized by maximal predictability for both particles: 
$P_{i,A}=P_{i,B} = 1$ and $V_{i,A}=V_{i,B} =C_i= 0$.
We observe that, as a consequence of the scattering, other CCR terms develop at expense of the predictability, 
according to Eq.\eqref{2}. This values change due to the scattering process, 
and in Fig.\ref{figone} we report both the non-relativistic case (Figs.\ref{fig1a}-\ref{fig1c}) 
and the relativistic one (Figs.\ref{fig1d}-\ref{fig1f}) characterized by different values of $\mu \equiv {|\bf{p}|}/{m_e}$, 
where $|\bf{p}|$ is the incoming momentum and $m_e$ the electron mass.  

As first important remark, we see that CCR as in Eq. \eqref{2} are verified after the scattering in any regimes. 
Equivalent results are obtained if one considers the entropic approach (Eqs.\eqref{0.5}-\eqref{0.6}). 
In particular, the general form expressed by Eq.\eqref{0.3}, is obtained if one identify the density matrix elements as
\begin{equation}
    \rho_{ij,kl} = \mathcal{M}(RL,ij)\mathcal{M}(RL,kl),
\end{equation}
where $i,j,k,l = R,L$ and $\mathcal{M}\in \mathbf{R}$.

Also we note that there is a symmetry between the two subsystems $A$ and $B$ 
for which predictability and visibility coincide. 
Moreover, comparing Fig.\ref{fig1c} and Fig.\ref{fig1f} we see that in the non-relativistic regime 
visibility develops, while it remains zero in the relativistic case. 

\subsection{General factorized states}

In this Section we consider a generalization of the previous case, 
in which the incoming particles are in a superposition of helicity states. 
We  have that the initial state is again a product state, 
but with predictability and visibility different from zero:
\begin{equation}
\label{instateII} 
    \ket{i}_{I\!I} = \big(\cos{\alpha}\ket{R}_A+ e^{i\xi}\sin{\alpha}\ket{L}_A\big)\otimes\big(\cos{\beta}\ket{R}_B+ e^{i\eta}\sin{\beta}\ket{L}_B\big).
\end{equation}
A particular case of such an initial state, obtained for $\al=0$ and $\xi=\eta=0$, 
has been studied in detail in Ref.\cite{Blasone:2024dud}.

The final state results
\begin{eqnarray}
\label{finstateII} \nonumber
    \ket{f}_{I\!I} &=&  \sum_{r,s}\Big[\cos{\alpha}\cos{\beta}\,\mathcal{M}(RR;rs)\ket{r}_A\ket{s}_B + e^{i\eta}\cos{\alpha}\sin{\beta}\, \mathcal{M}(RL;rs)\ket{r}_A\ket{s}_B + e^{i\xi}\sin{\alpha}\cos{\beta}\,\mathcal{M}(LR;rs)\ket{r}_A\ket{s}_B \\
      && \qquad \qquad \qquad +\,  e^{i(\xi+\eta)}\sin{\alpha}\sin{\beta}\,\mathcal{M}(LL;rs)\ket{r}_A\ket{s}_B\Big].
\end{eqnarray}
\begin{figure}[!]
 \subfloat[][\emph{}]{\includegraphics[width =5.3cm]{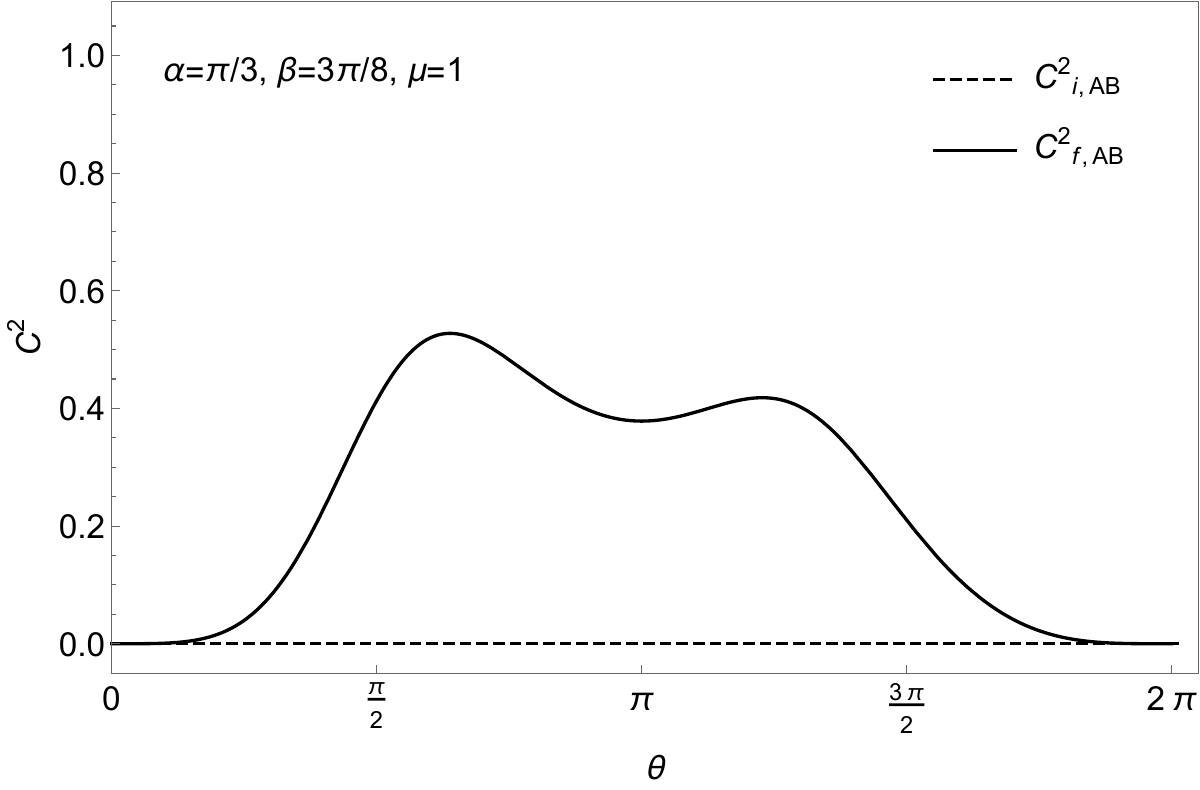}
\label{fig2a}}\quad
 \subfloat[][\emph{}]{\includegraphics[width =5.55 cm]
 {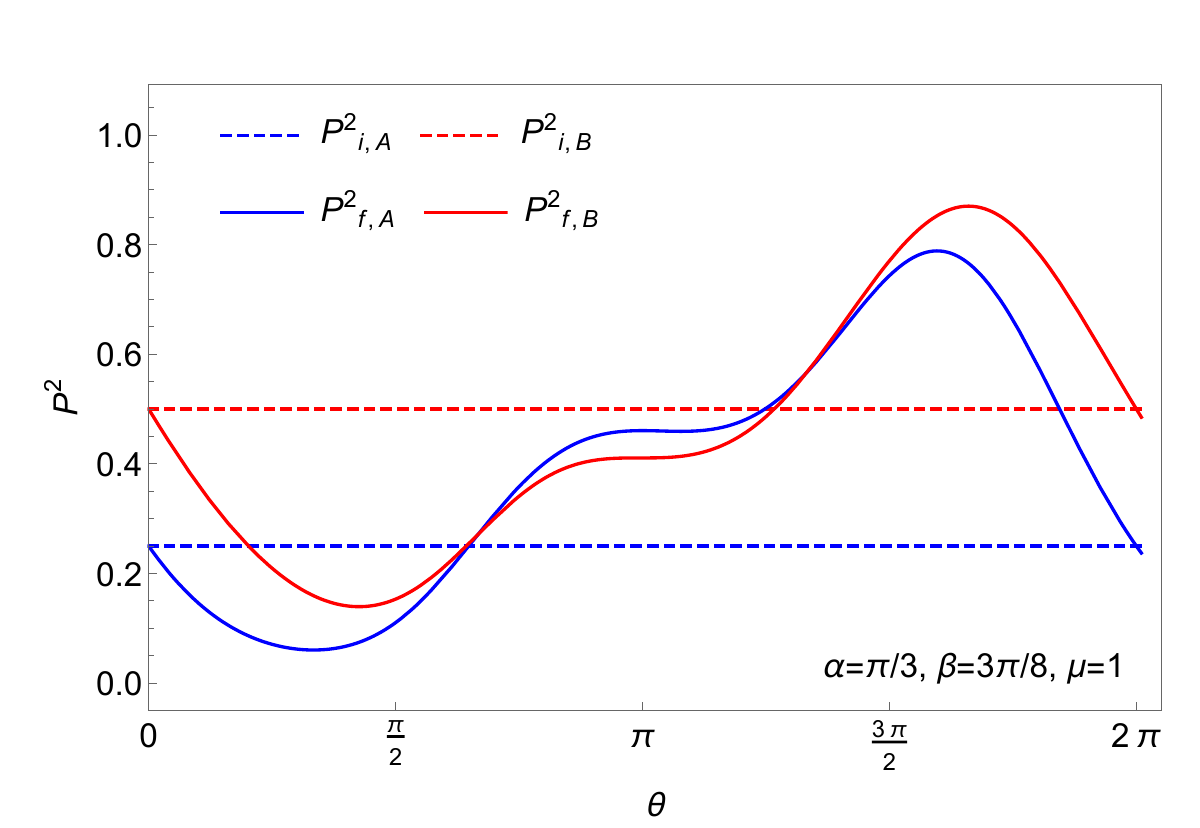}
\label{fig2b}}\quad
 \subfloat[][\emph{}]{\includegraphics[width =5.55 cm]{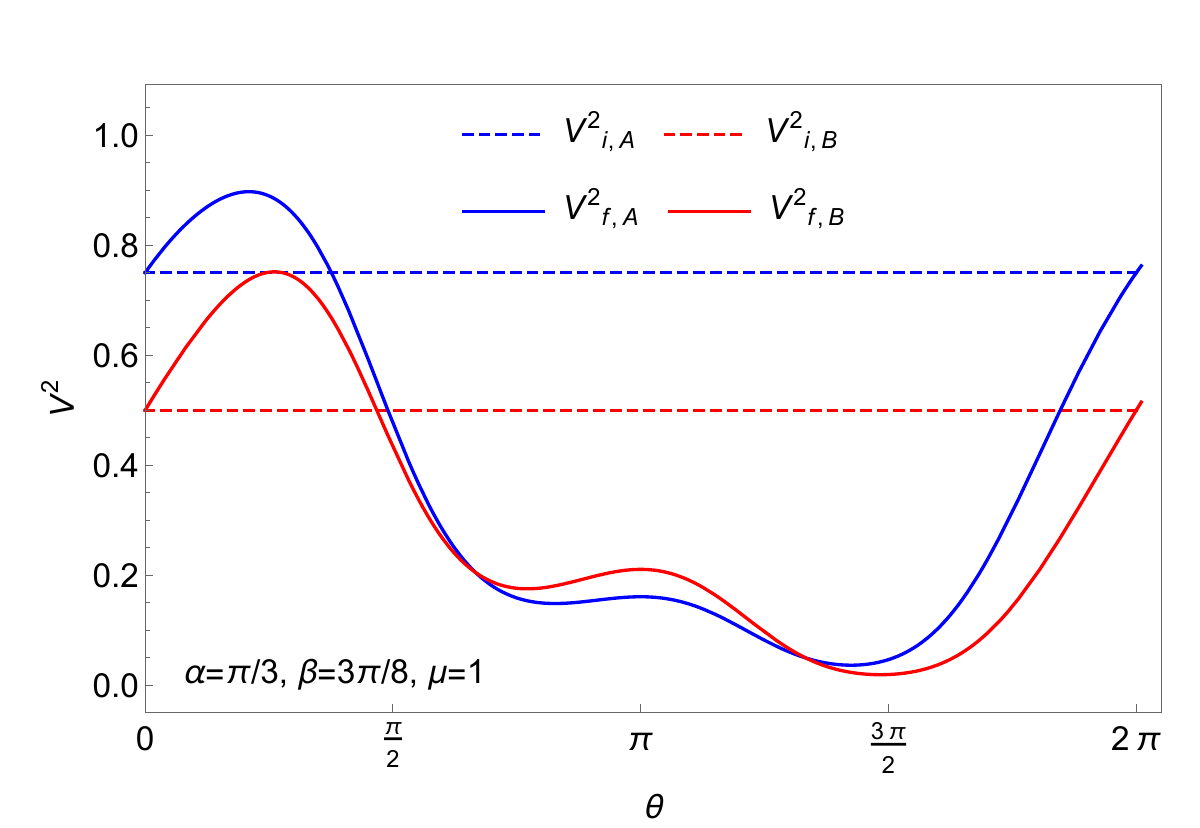}
\label{fig2c}}\quad
 \subfloat[][\emph{}]{\includegraphics[width =5.3 cm]{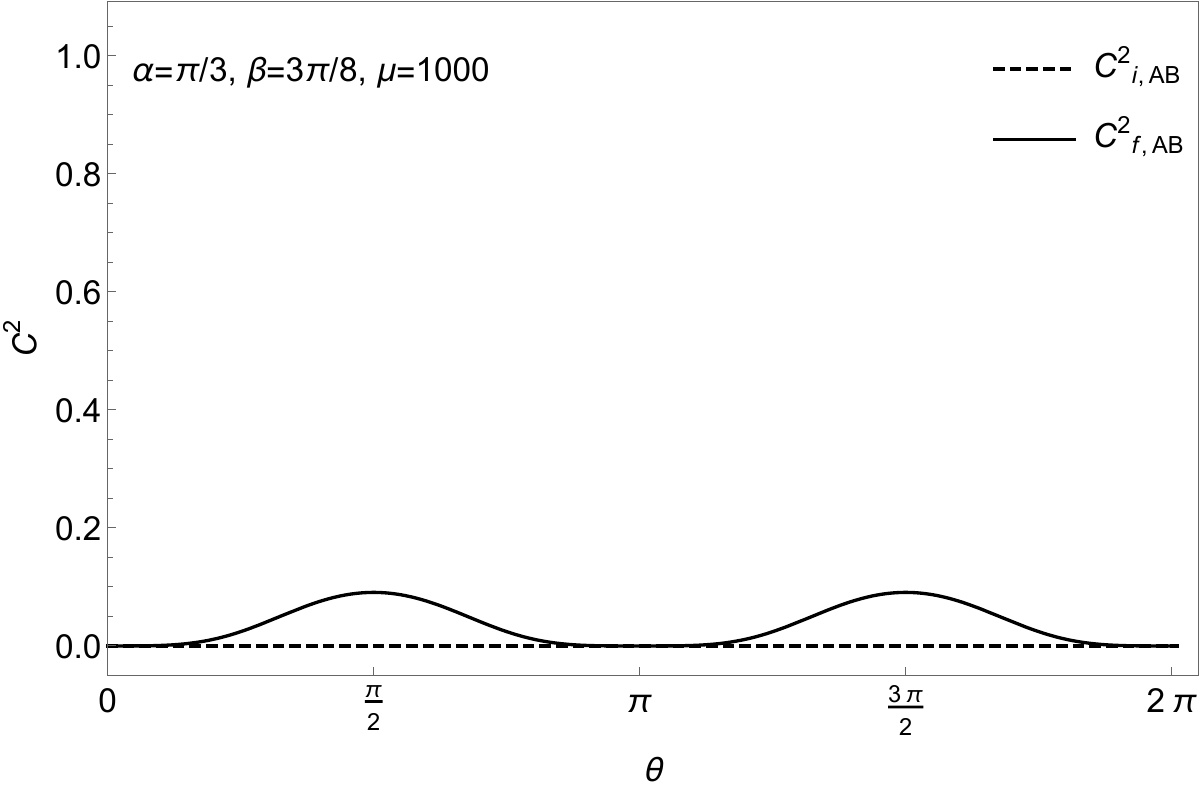}
\label{fig2d}}\quad
 \subfloat[][\emph{}]{\includegraphics[width =5.55 cm]{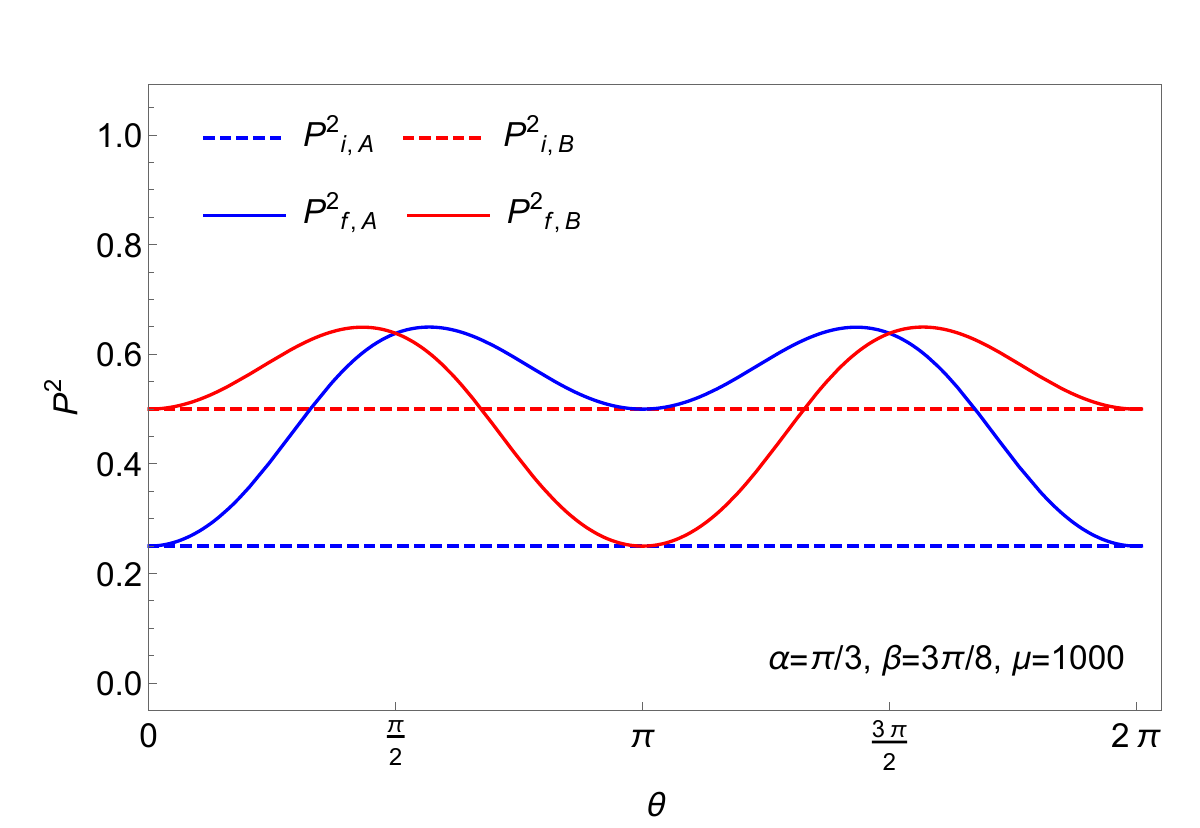}
\label{fig2e}}\quad
 \subfloat[][\emph{}]{\includegraphics[width =5.55 cm]{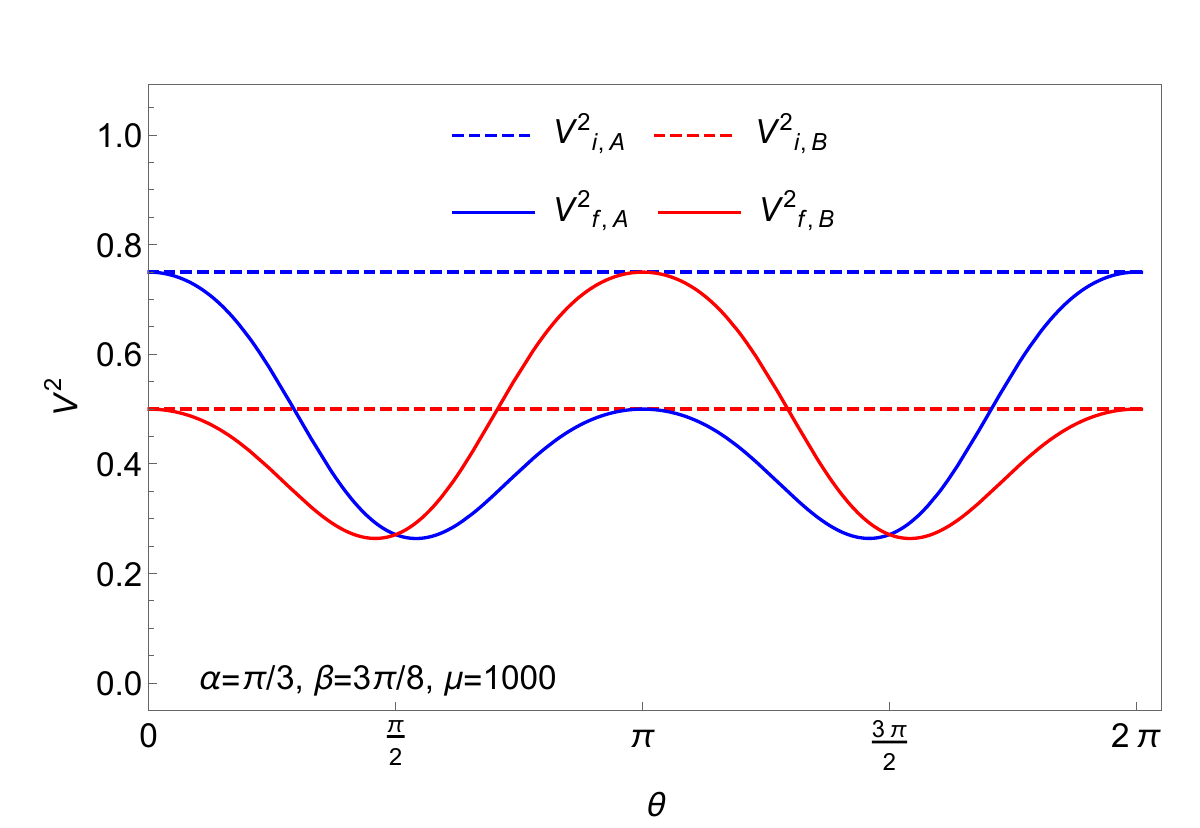}
\label{fig2f}}\quad
\caption{Plot of the CCR terms $C^2$, $P^2$ and $V^2$, for initial and final states of Case II.
Panels (a),(b),(c) refer to the value $\mu=1$ (non-relativistic regime). 
Panels (d),(e),(f) refer to the value $\mu=1000$ (relativistic regime).} 
\label{figtwo}
\end{figure}

For simplicity, we set $\eta=\xi=0$ and refer to Figs.\ref{fig2a}-\ref{fig2c} for the non-relativistic case, 
and to Figs.\ref{fig2d}-\ref{fig2f} for the relativistic one.
It is simple to verify that Eq.\eqref{2} holds again. 
At variance with  case I, in the non-relativistic regime 
we note that the symmetry between $A$ and $B$ is lost due to the asymmetry already present in the initial state, 
and the two parts show different predictability and visibility. 
Furthermore in this case, an asymmetry emerges  with respect to the scattering angle $\theta=\pi$. 
It can be verified that this is a subtle consequence of the off-diagonal terms contributions. 
\smallskip

In the relativistic case, symmetry with respect to $\theta=\pi$ is restored, 
and visibility is non-zero simply because it is already non vanishing before the scattering.
As in the previous case, we can verify Eq.\eqref{0.5}, by promoting the matrix elements to the form
\begin{equation}
    \rho_{i,j,k,l} = A_{\zeta,\kappa}A^{*}_{\zeta',\kappa'}\mathcal{M}(i,j,\zeta,\kappa)\mathcal{M} (k,l,\zeta',\kappa'),
    \label{rho2}
\end{equation}
where the repeated indices $\zeta,\kappa = R,L$ are summed 
according to the Einstein convention and 
$A_{R,R} = \cos{\alpha}\cos{\beta}$, $A_{R,L} = e^{i\eta}\cos{\alpha}\sin{\beta}$, 
$A_{L,R} = e^{i\xi}\sin{\alpha}\cos{\beta}$ and 
$A_{L,L} = e^{i(\eta+\xi)}\sin{\alpha}\sin{\beta}$. The form (\ref{rho2}) is
the most general one, and remains the same also in the next case C.

It is interesting to consider the relativistic limit $\mu=\infty$ for which it is possible to obtain analytical forms of the CCR quantities for the state in Eq.\eqref{finstateII}:
\begin{equation}
\lim_{\mu\rightarrow\infty}C = \frac{4}{D_{I\!I}}\Big(1-8\cos{[2(\alpha-\beta)]}+7\cos{[2(\alpha+\beta)]}-2\cos{2\theta}\sin^2{[\alpha+\beta]}\Big)\sin^2{\theta},
\end{equation}
\begin{equation}
   \lim_{\mu\rightarrow\infty}P_A  = \frac{8}{D_{I\!I}}\Big[\cos{2\beta}(\cos{3\theta}+7\cos{\theta}-8)-\cos{2\alpha}(\cos{3\theta}+7\cos{\theta}+8)\Big], 
\end{equation}
\begin{equation}
   \lim_{\mu\rightarrow\infty}P_B  = \frac{8}{D_{I\!I}}\Big[\cos{2\alpha}(\cos{3\theta}+7\cos{\theta}-8)-\cos{2\beta}(\cos{3\theta}+7\cos{\theta}+8)\Big], 
\end{equation}
\begin{equation}
   \lim_{\mu\rightarrow\infty}V_A = \frac{64}{D_{I\!I}}\left(\cot^4{\frac{\theta}{2}}\sin{2\alpha}+\sin{2\beta}\right)\sin^4{\frac{\theta}{2}}, 
\end{equation}
\begin{equation}
   \lim_{\mu\rightarrow\infty}V_B = \frac{64}{D_{I\!I}}\left(\cot^4{\frac{\theta}{2}}\sin{2\beta}+\sin{2\alpha}\right)\sin^4{\frac{\theta}{2}}. 
\end{equation}
with
\be
D_{I\!I}\equiv  2(7+\cos^2{2\theta})^2+4\cos{2\alpha}\cos{2\beta}(15+\cos{2\theta})\sin^2{\theta}+8\sin{2\alpha}\sin{2\beta}\sin^4{\theta}
\ee
As it can be easily checked, the sum of the squares of this three quantities is exactly one, 
for each subsystem, according to Eq.\eqref{2}.

\subsection{General incoming state}
\label{sec3c}

Finally we consider the case of the most general initial state, 
parameterized by six real parameters (three angles and three phases), 
and configurations in which entanglement is present from the beginning: 
\begin{equation}
\label{instategen} 
    \ket{i}_{I\!I\!I} = \cos{\alpha}\ket{R}_A\ket{R}_B + 
		e^{i\xi}\sin{\alpha}\cos{\beta}\ket{R}_A\ket{L}_B + 
		e^{i\eta}\sin{\alpha}\sin{\beta}\cos{\chi}\ket{L}_A\ket{R}_B + 
		e^{i\tau}\sin{\alpha}\sin{\beta}\sin{\chi}\ket{L}_A\ket{L}_B.
\end{equation}

\smallskip 

The corresponding final state is given by
\begin{eqnarray}
\label{finstategen} \nonumber
    \ket{f}_{I\!I\!I} &=&  \sum_{r,s}\Big[\cos{\alpha}\,\mathcal{M}(RR;rs)\ket{r}_A\ket{s}_B + 
		e^{i\xi}\sin{\alpha}\cos{\beta}\,\mathcal{M}(RL;rs)\ket{r}_A\ket{s}_B + 
		e^{i\eta}\sin{\alpha}\sin{\beta}\cos{\chi}\,\mathcal{M}(LR;rs)\ket{r}_A\ket{s}_B \\
      && \qquad \qquad \qquad +\,  e^{i\tau}\sin{\alpha}\sin{\beta}\sin{\chi}\, \mathcal{M}(LL;rs)\ket{r}_A\ket{s}_B\Big].
\end{eqnarray}

The matrix elements $\rho_{i,j,k,l}$ have the same form of Eq.\eqref{rho2} 
with $A_{R,R} = \cos{\alpha}$, 
$A_{R,L} = e^{i\xi}\sin{\alpha}\cos{\beta}$, $A_{L,R} = 
e^{i\eta}\sin{\alpha}\sin{\beta}\cos{\chi}$ 
and $A_{L,L} = e^{i\tau}\sin{\alpha}\sin{\beta}\sin{\chi}$.

\smallskip

In the following we show that, for some choice of parameters, 
the entanglement resulting after the scattering process is in a non-trivial relation with respect to its initial value. 
We illustrate this by providing some significant examples.
Consider the two configurations $\mathcal{A} = \bigl(\alpha = \pi/4, \beta = \pi/6, 
\chi = \pi/2\bigr)$ and $\mathcal{B} = \bigl(\alpha = \pi/3, \beta = \pi/6, \chi = 0\bigr)$.

\begin{figure}[t]
\subfloat[][\emph{}]{\includegraphics[width =5.3 cm]{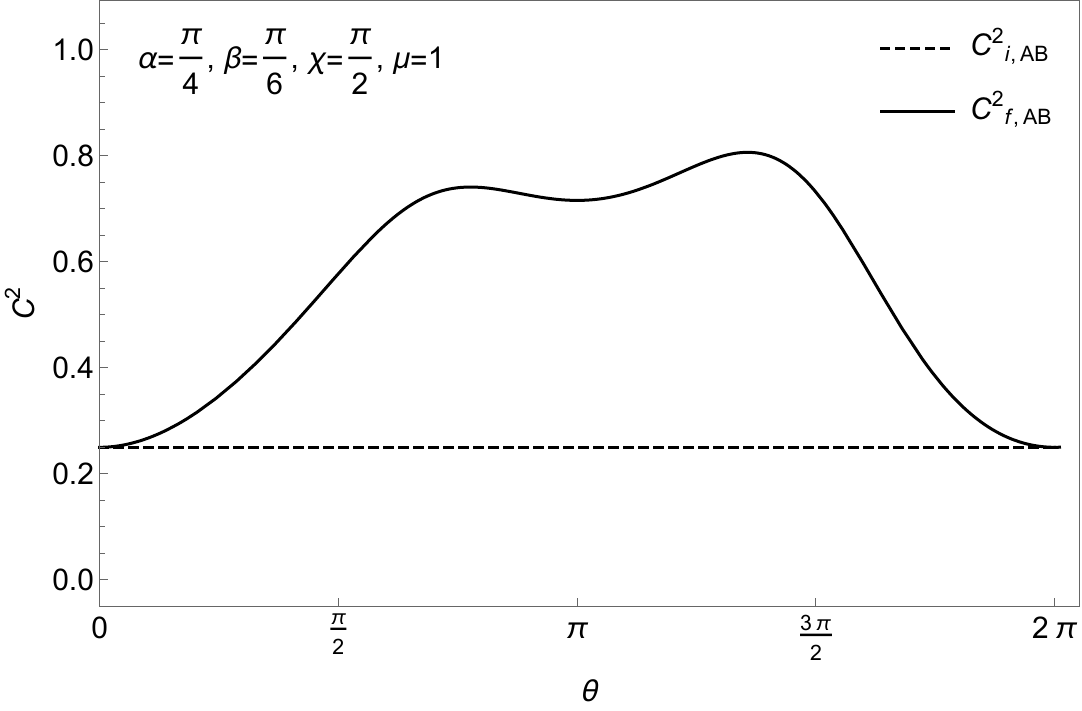}
\label{fig3a}}\quad
 \subfloat[][\emph{}]{\includegraphics[width =5.55 cm]{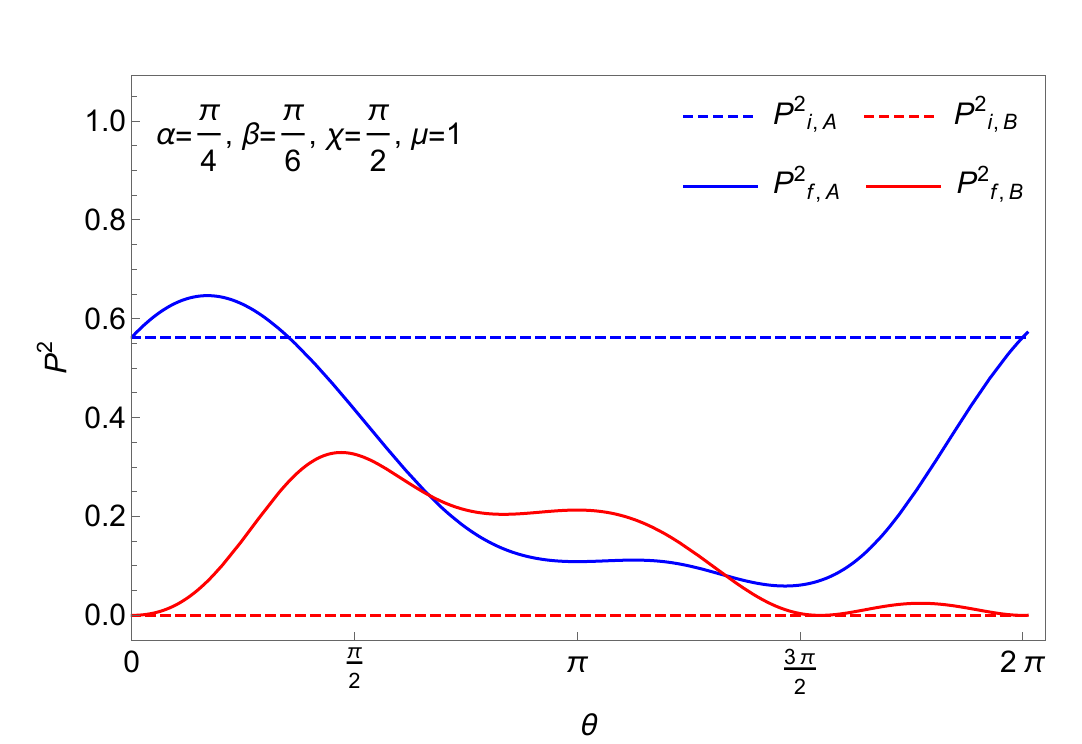}
\label{fig3b}}\quad
 \subfloat[][\emph{}]{\includegraphics[width =5.55 cm]{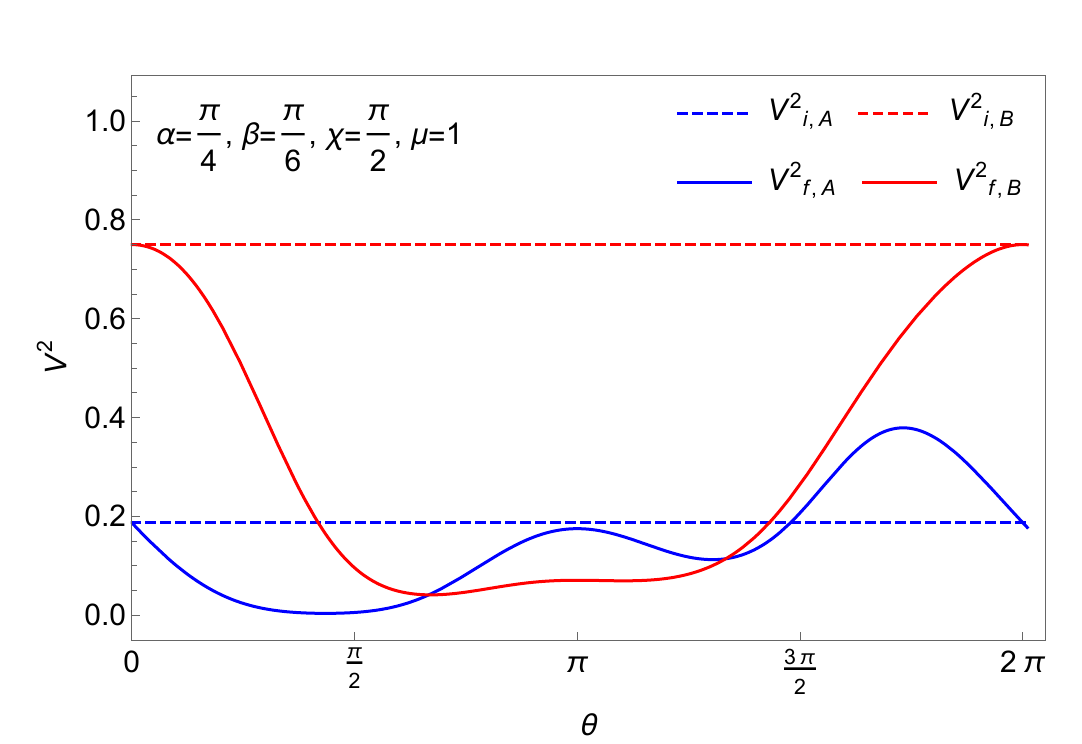}
\label{fig3c}}\quad
\subfloat[][\emph{}]{\includegraphics[width =5.3 cm]{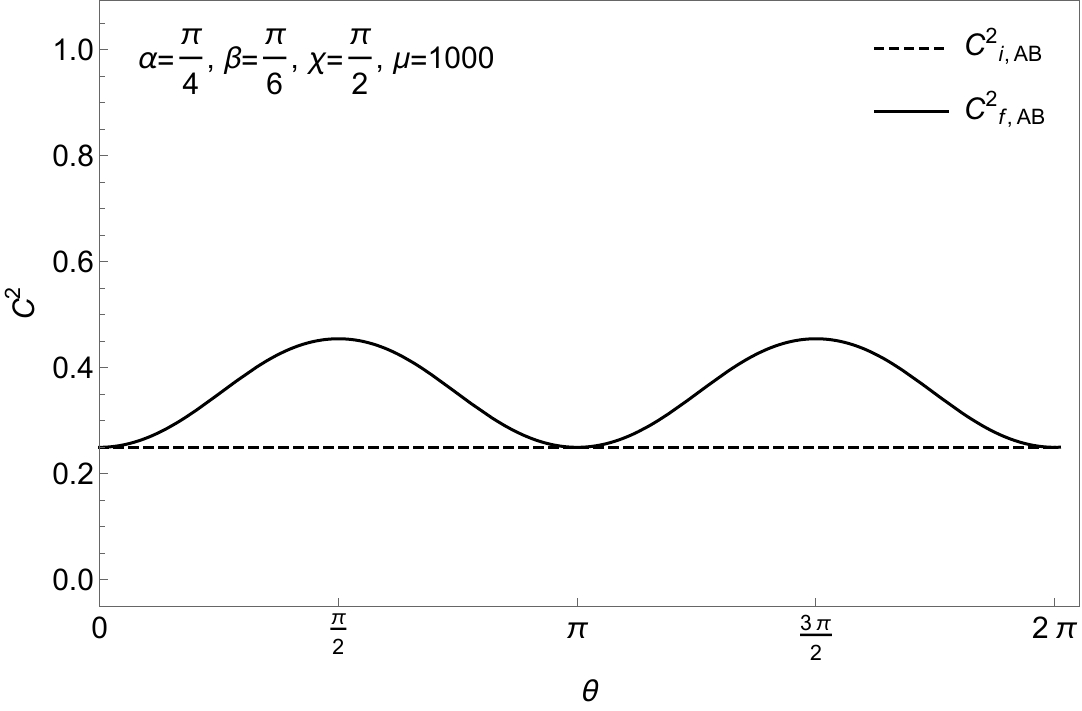}
\label{fig3d}}\quad
 \subfloat[][\emph{}]{\includegraphics[width =5.55 cm]{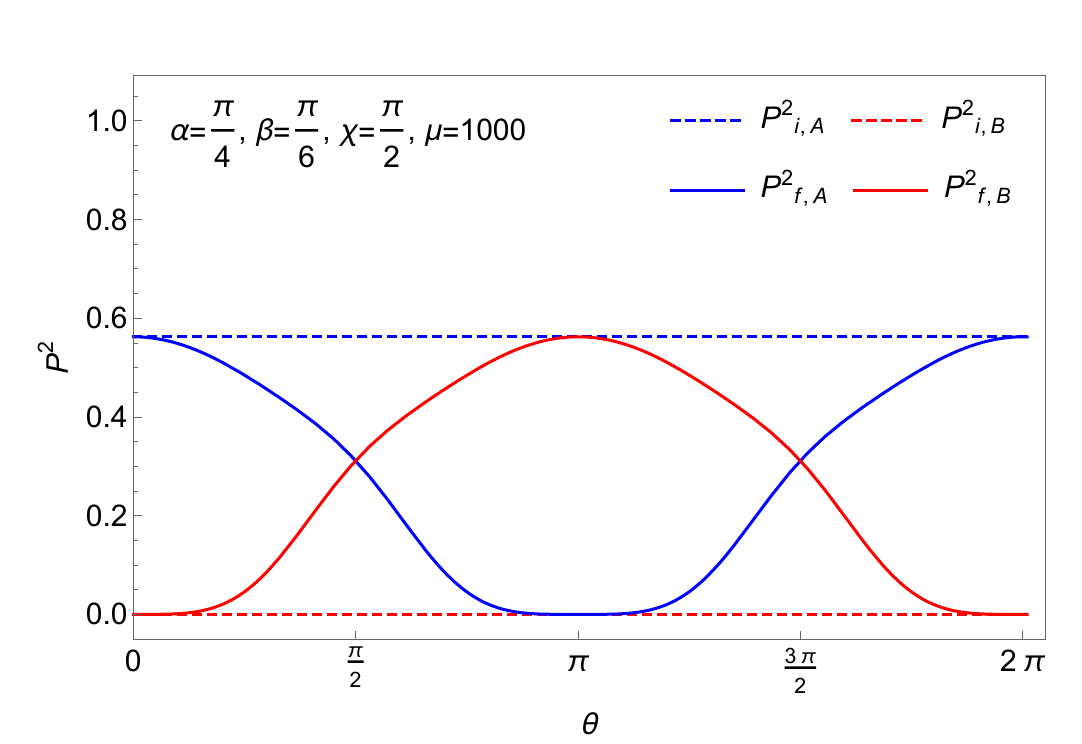}
\label{fig3e}}\quad
\subfloat[][\emph{}]{\includegraphics[width =5.55 cm]{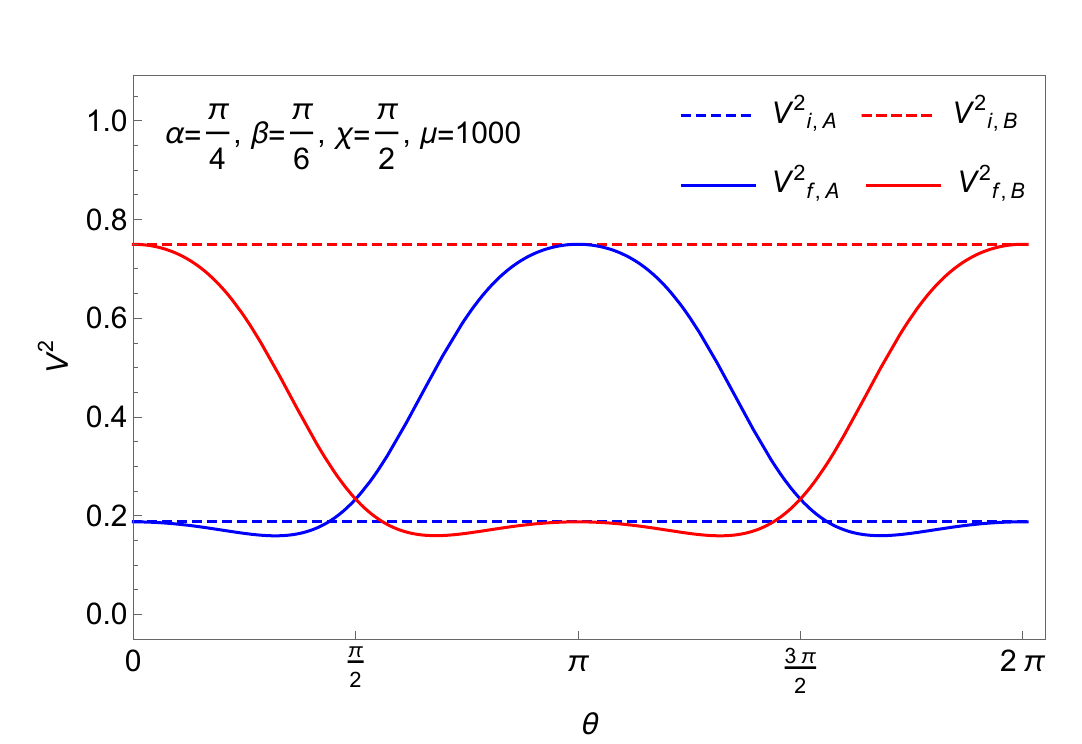}
\label{fig3f}}\quad
\caption{Plot of CCR terms for Case III with low value of initial entanglement.} 
\label{figthree}
\end{figure}

\begin{figure}[t]
 \subfloat[][\emph{}]{\includegraphics[width =5.3 cm]{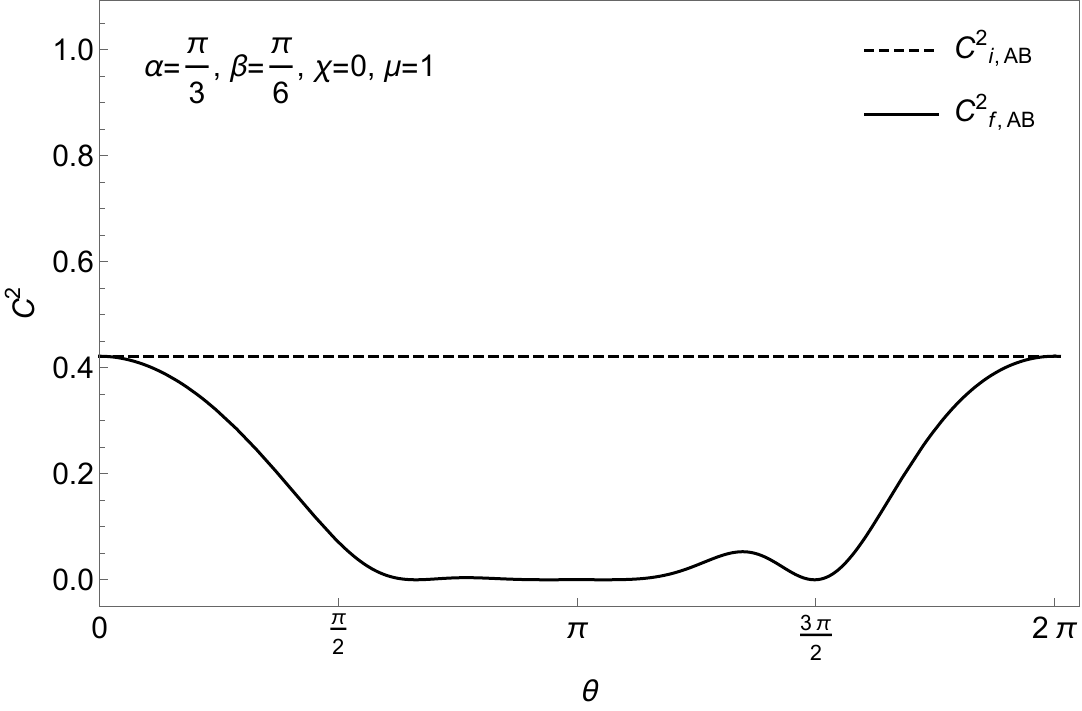}
\label{fig4a}}\quad
 \subfloat[][\emph{}]{\includegraphics[width =5.55 cm]{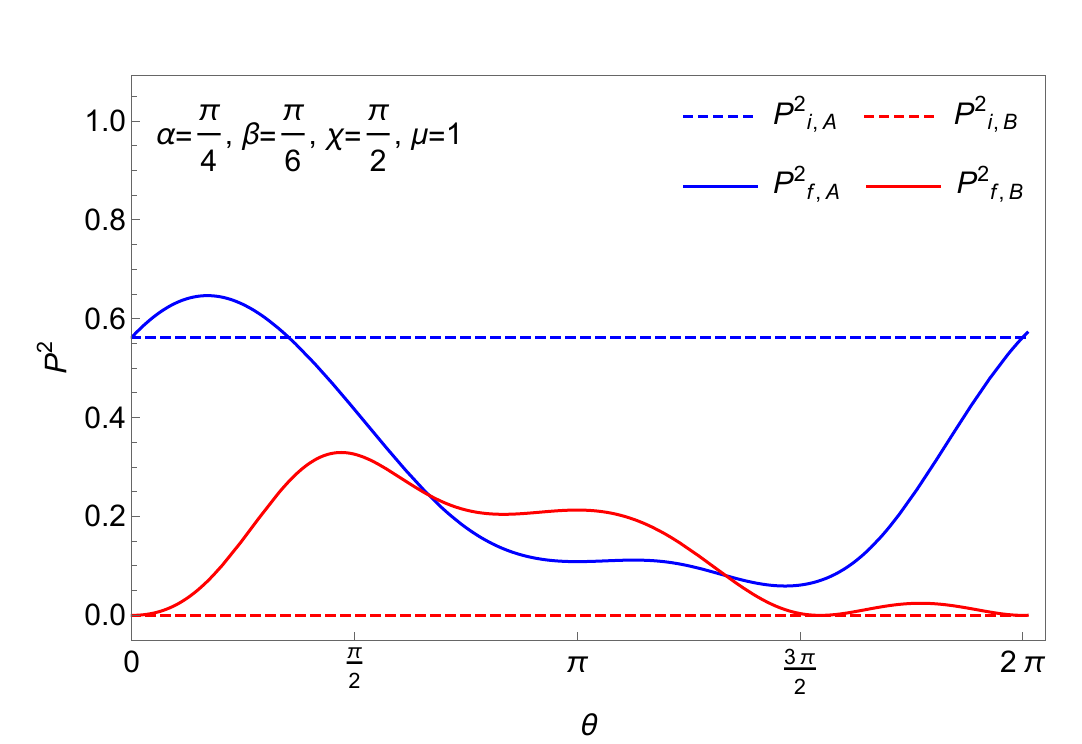}
\label{fig4b}}\quad
 \subfloat[][\emph{}]{\includegraphics[width =5.55 cm]{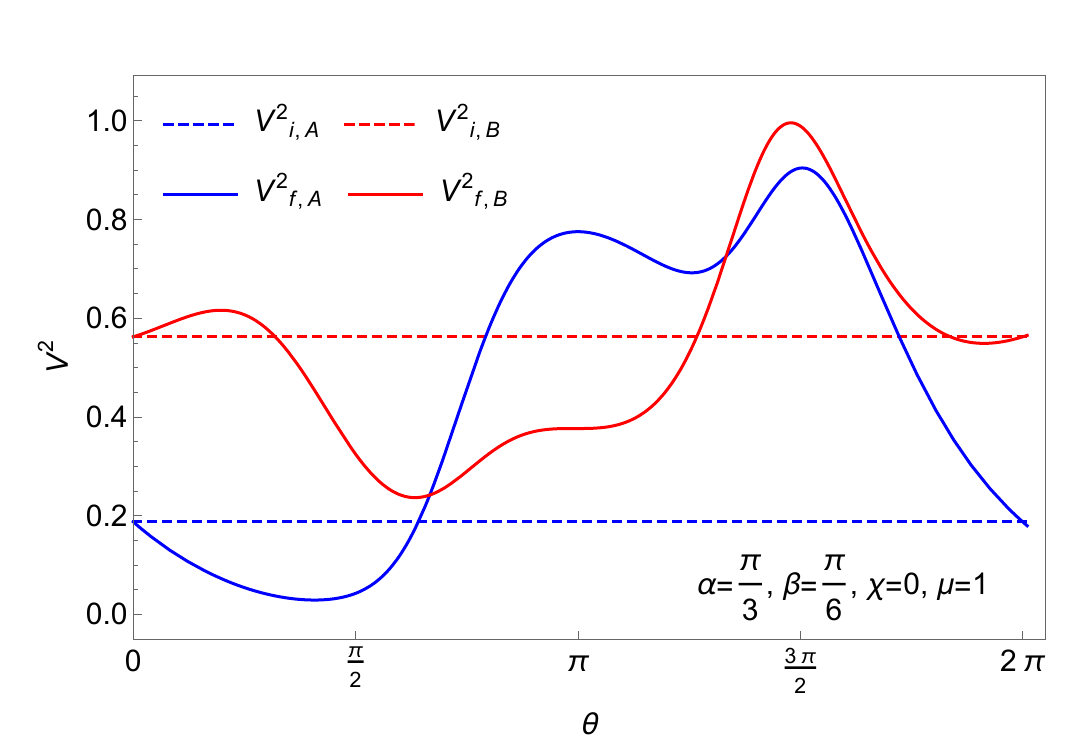}
\label{fig4c}}\quad
 \subfloat[][\emph{}]{\includegraphics[width =5.3 cm]{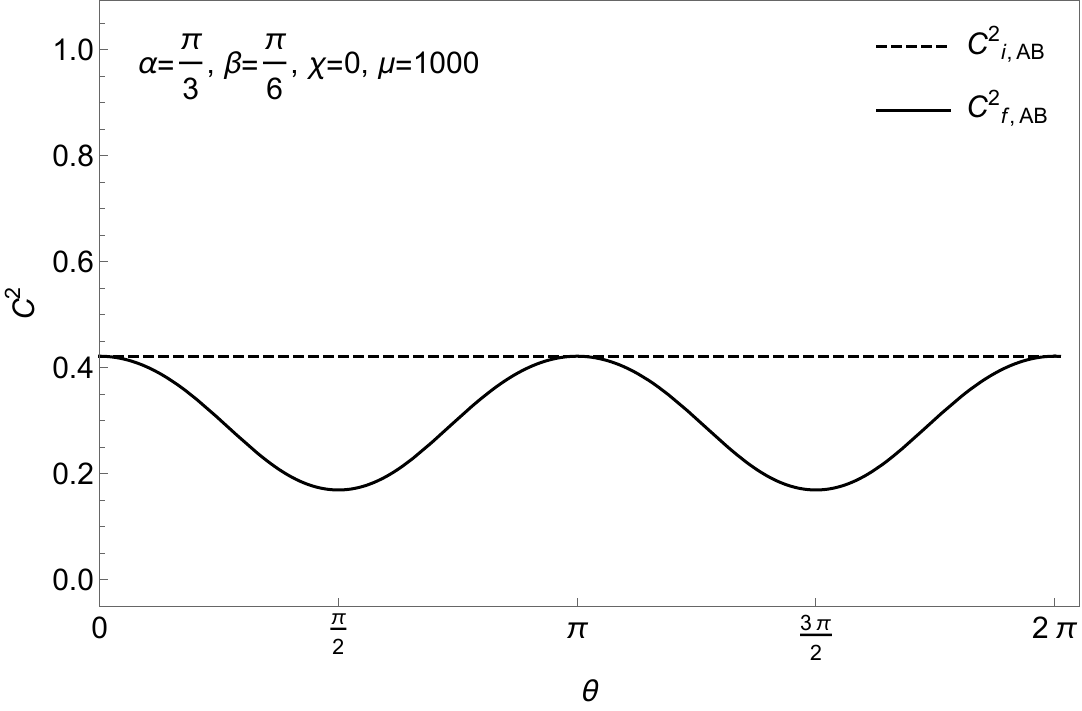}
\label{fig4d}}\quad
 \subfloat[][\emph{}]{\includegraphics[width =5.55 cm]{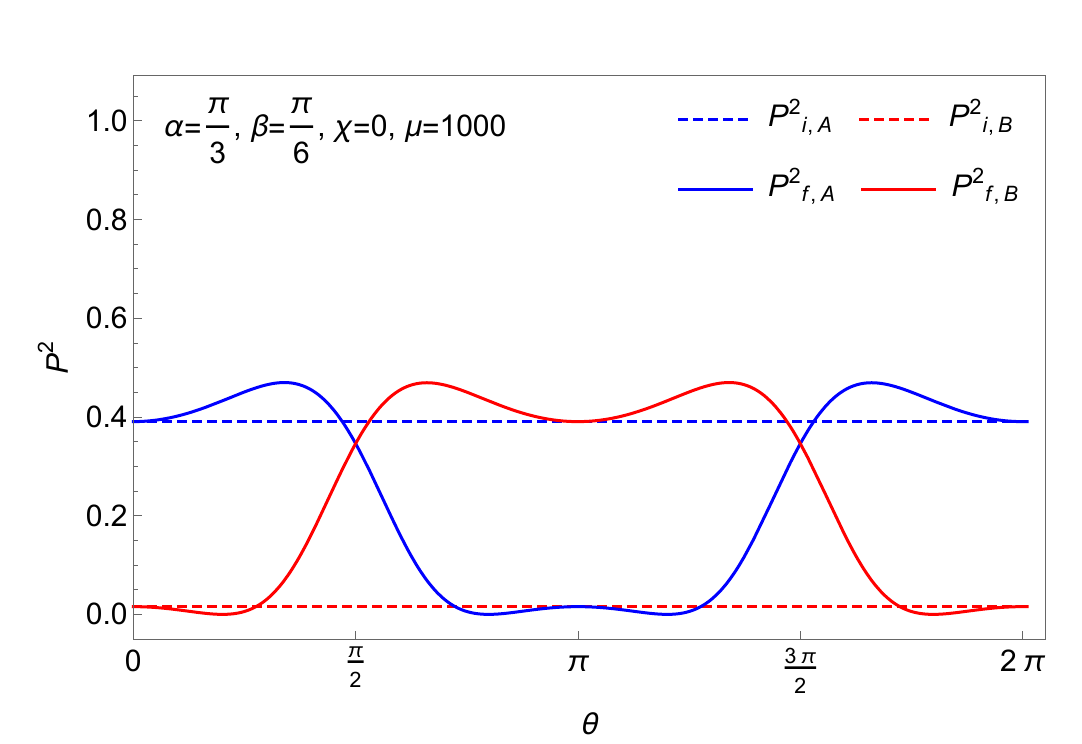}
\label{fig4e}}\quad
 \subfloat[][\emph{}]{\includegraphics[width =5.55 cm]{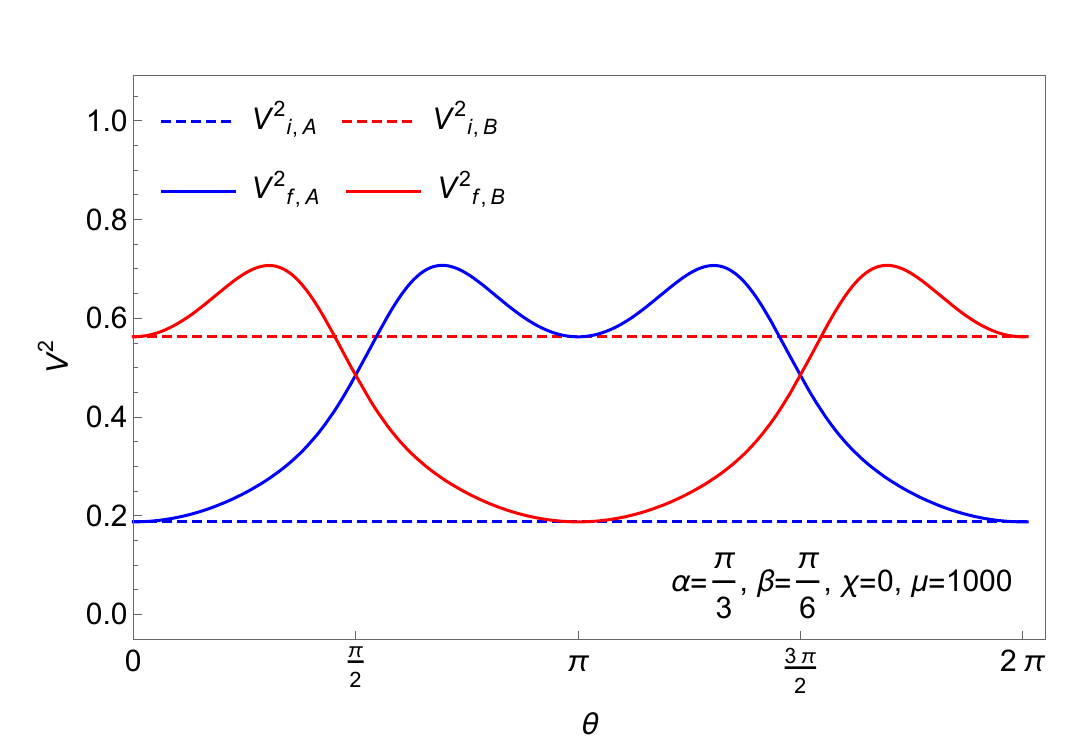}
\label{fig4f}}\quad
\caption{Plot of CCR terms for Case III with  nonzero initial value for all three quantities.} 
\label{figfour}
\end{figure}

\begin{figure}[h]
 \subfloat[][\emph{}]{\includegraphics[width =5.55 cm]{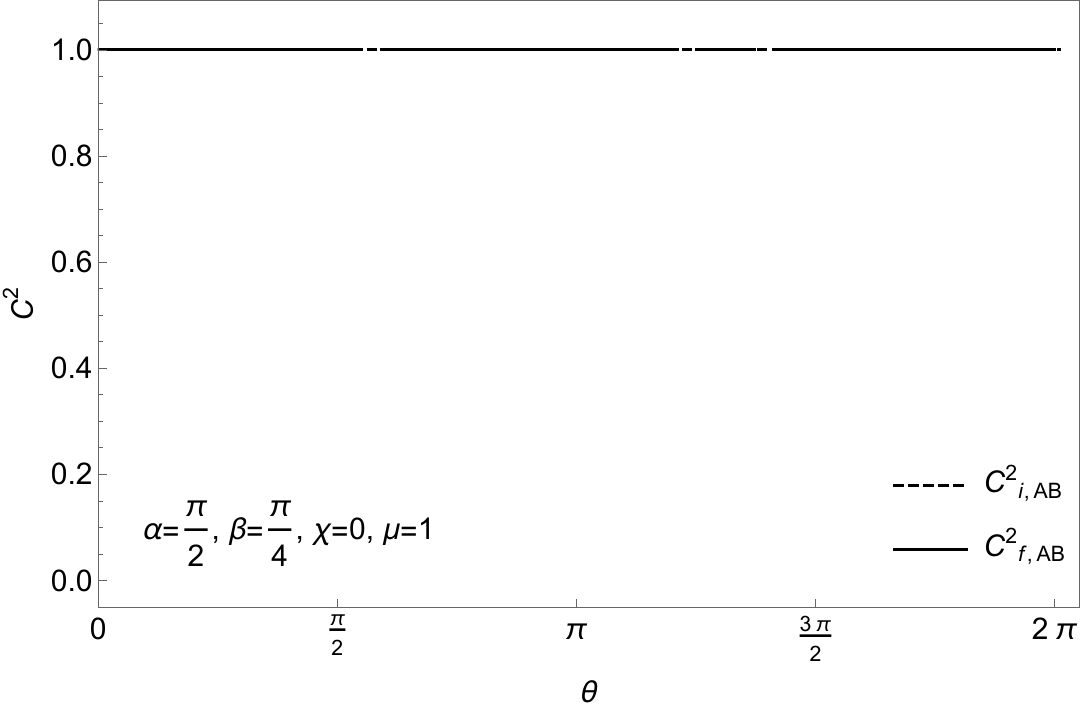}
\label{fig5a}}\quad
 \subfloat[][\emph{}]{\includegraphics[width =5.55 cm]{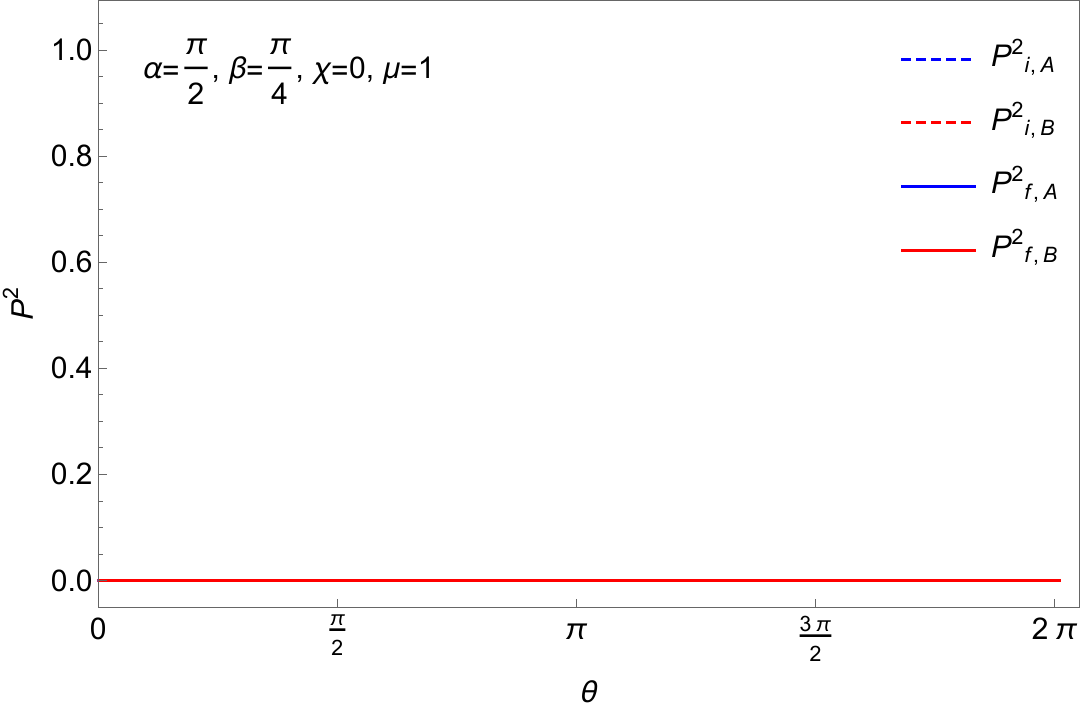}
\label{fig5b}}\quad
 \subfloat[][\emph{}]{\includegraphics[width =5.55 cm]{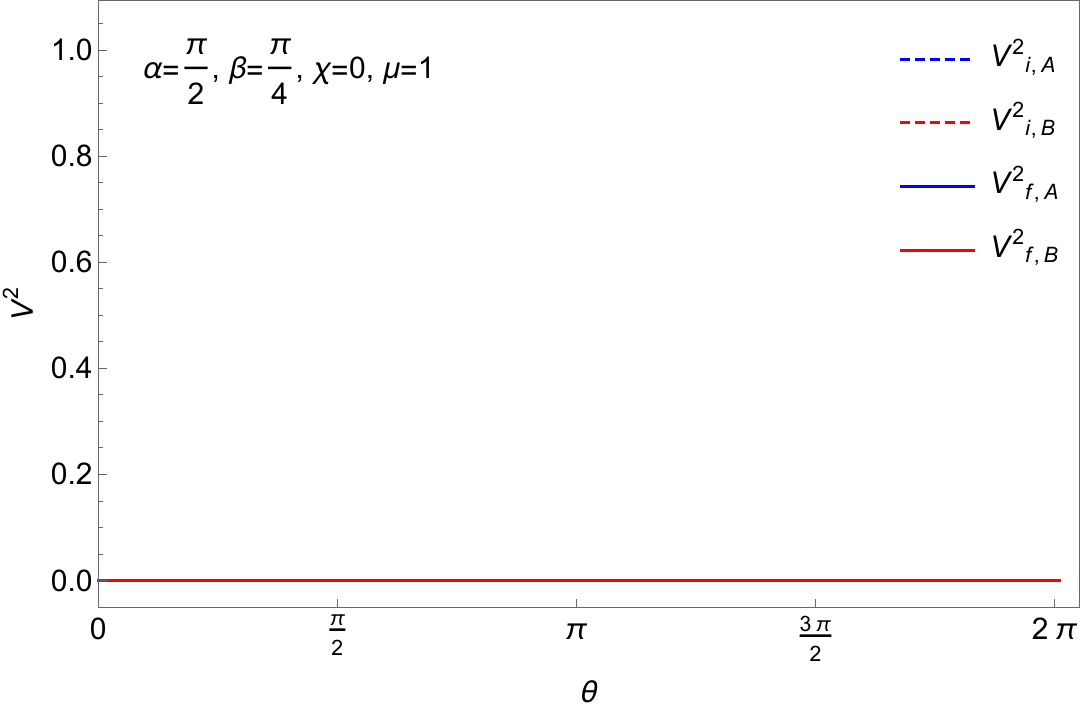}
\label{fig5c}}\quad
\caption{Plot of CCR terms for Case III with initial Bell state.} 
\label{figfive}
\end{figure}

For the configuration $\mathcal{A}$, in
Figs.\ref{fig3a}-\ref{fig3c} we plot the non-relativistic case, 
while Figs.\ref{fig3d}-\ref{fig3f} refer to the relativistic one. 
From Figs.\ref{fig3a} and \ref{fig3d} we can see that  entanglement increases in both non-relativistic and relativistic regimes. 

On the other hand, when configuration $\mathcal{B}$ is adopted (see Fig.\ref{figfour}), 
we observe the opposite behaviour. In particular, in the non-relativistic regime, the final entanglement is strongly suppressed and for some values of $\theta$ it vanishes.



\section{Classification of initial states by entanglement behaviour under scattering}

The non-trivial behaviour of entanglement in the scattering processes, observed in the previous section, suggests to search for a classification of states with similar properties in this respect. In the following we show that one can identifies three classes of states which we define as \textit{entanglophilus}, \textit{entanglophobous} and \textit{mixed}. Furthermore, a very interesting situation arises when the initial state is  maximally entangled.

\subsection{Maximally entangled states}
\label{MES}

Following the discussion of Section \ref{sec3c}, we consider as initial state the Bell state
$\Psi^+ = (\ket{RL}+\ket{LR})/\sqrt{2}$: from Fig.\ref{figfive}, we see that the initial maximal entanglement is conserved at any $\theta$ after the scattering process and, consistently, no visibility and predictability arise. This very remarkable behaviour induces us to investigate the case of generic maximally entangled initial states for Bhabha scattering and also for other QED processes.
We consider separately four processes, in which only spin-$1/2$ massive particles are involved, and other two in which also spin-$1$ massless particles, as in- and out-states, are present (see tables below). 
By means of the scattering amplitudes reported in Appendix we have verified that for the first four processes there is a complete conservation of maximum entanglement: 
if the input state is any maximally entangled state, 
then the final state is maximally entangled too, 
whatever the scattering angle and the value of $\mu$
(\textit{complete maximal entanglement conservation}). 
At variance, for the last two processes this is in general not true:
for the process of annihilation $e^-e^+\rightarrow \gamma\gamma$ maximal concurrence conservation partially survives, while
for the Compton process there is no conservation of maximal entanglement. 

For any process it is clarifying to consider, as initial maximally entangled states, the four Bell states 
\begin{equation}
\Phi^{\pm} = \frac{1}{\sqrt{2}} (\ket{RR} \pm \ket{LL}), \qquad
\Psi^{\pm} = \frac{1}{\sqrt{2}} (\ket{RL} \pm \ket{LR}).
\label{BellStates}
\end{equation}
with the corresponding final states.
In the tables below we report the correspondences for each process:
\begin{table}[h]
    \begin{tabular}{|c|c|c|c|c|c|c|}
\hline
 & Bhabha & &  & & M\o{}ller & \\
\hline
In. State & Fin. State & Conc. &  & In. State & Fin. State & Conc. \\ 

\hline
$\Phi^{+}$ & $\Phi^{+}$ & $1$ &  & $\Phi^{+}$ & $\cos{s_3} \, \Phi^{+} + \sin{s_3} \, \Psi^{-}$ & $1$ \\
\hline
$\Phi^{-}$ & $\cos{s_1} \, \Phi^{-} + \sin{s_1} \, \Psi^{+}$ & $1$ &  & $\Phi^{-}$ & $\Phi^{-}$ & $1$ \\
\hline
$\Psi^{+}$ & $\cos{s_2} \, \Phi^{-} + \sin{s_2} \, \Psi^{+}$ & $1$ &  & $\Psi^{+}$ & $\Psi^{+}$ & $1$ \\
\hline
$\Psi^{-}$ & $\Psi^{-}$ & $1$ &  & $\Psi^{-}$ & $\cos{s_4} \Phi^{+} + \sin{s_4} \, \Psi^{-}$ & $1$ \\
\hline
\end{tabular}
    \caption{Bell states transformation and entanglement of final states for Bhabha and M\o{}ller scattering processes.}
    \label{tab1}
\end{table}

\vspace{0.5cm}

\begin{table}[t]
    \begin{tabular}{|c|c|c|c|c|c|c|}
\hline
 & $e^{-} e^{+} \rightarrow \mu^{-}\mu^{+}$ & &  &  & $e^{-} \mu^{-} \rightarrow e^{-} \mu^{-}$ & \\
\hline
In. State & Fin. State & Conc. &  & In. State & Fin. State & Conc. \\ 
\hline
$\Phi^{+}$ & $\Phi^+$  & $1$ &  & $\Phi^{+}$ & $\cos{s_7} \, \Phi^{+} + \sin{s_7} \, \Psi{-}$ & $1$ \\
\hline
$\Phi^{-}$ & $\cos{s_5} \Phi^{+} + \sin{s_5} \, \Psi^{-}$  & $1$ &  & $\Phi^{-}$ & $\cos{s_8} \, \Phi^{-} + \sin{s_8} \, \Psi{+}$ & $1$ \\
\hline
$\Psi^{+}$ & $\cos{s_6} \, \Phi^{+} + \sin{s_6} \, \Psi^{-}$& $1$ &  & $\Psi^{+}$ & $\cos{s_9} \, \Phi^{-} + \sin{s_9} \, \Psi{+}$ & $1$ \\
\hline
$\Psi^{-}$ & $\Psi^{-}$& $1$ &  & $\Psi^{-}$ & $\cos{s_{10}} \, \Phi^{+} + \sin{s_{10}} \, \Psi{-}$ & $1$ \\
\hline
\end{tabular}
    \caption{Bell states transformation and entanglement of final states for electron-positron annihilation in two muons (see footnote [83]) and electron-muon in electron-muon scattering processes.}
    \label{tab2}
\end{table}

\vspace{0.5cm}

\begin{table}[t]
    \begin{tabular}{|c|c|c|c|c|c|c|}
\hline
 & $e^{-} e^{+} \rightarrow \gamma\gamma$ & & & & Compton & \\
\hline
In. State & Fin. State & Conc. &  & In. State & Fin. State & Conc. \\ 
\hline
$\Phi^{+}$ & $\Phi{-}$ & $1$ &  & $\Phi^{+}$ & GC & $< 1$ \\
\hline
$\Phi^{-}$ & $\cos{r} \, \Phi^{+} + \sin{r} \, \Psi^{+}$ & $|\sin{(2 \, r)|}$ &  & $\Phi^{-}$ & GC & $< 1$ \\
\hline
$\Psi^{+}$ & $\Psi^{+}$ & $1$ &  & $\Psi^{+}$ & GC & $< 1$ \\
\hline
$\Psi^{-}$ & $\Psi^{-}$ & $1$ &  & $\Psi^{-}$ & GC & $< 1$ \\
\hline
\end{tabular}
        \caption{Bell states transformation and entanglement of final states for electron-positron annihilation in two photons and Compton scattering processes.}
    \label{tab3}
\end{table}

\vspace{0.5cm}

\noindent In the last table, referring to the Compton scattering,
``GC'' (Generic Coefficients) indicates that the coefficients of the four terms $\ket{rs}$, in which $r, s = R, L$, are all different, preventing a combination of Bell states with maximal concurrence.
The angles $\{s^i\}$ are suitably expressed in terms of the scattering amplitudes, for example, in the case of initial state $\Phi^{-}$
in the Bhabha scattering, $s_1 = \arccos{[\mathcal{N}^{- 1} \, (\mathcal{M} (RR; RR) - \mathcal{M} (RR; LL))]}$,
where $\mathcal{N} = \sqrt{(\mathcal{M} (RR; RR) - \mathcal{M} (RR; LL))^2 + 4 \, \mathcal{M} (RR; LL)^2}$ is the normalization factor.
Similarly for the other cases.

We see that the tables provide a sort of characterization of each process. 
The first four scattering processes convert the initial Bell states in themselves, 
or in a convex combination of Bell states again with concurrence one, 
which keep the same superposition form if the process is repeated \footnote{The achievement of a final, no more changed, combination with maximum concurrence is the common mechanism. For example, if as initial states one chooses the generalized Bell states
with a relative phase different from $0, \pi$, the final combination of concurrence $1$ is achieved after two iterations.},\footnote{For the process of annihilation in muons we recall that it can be realized only in ultrarelativistic regime, and thus only Bell states $\Psi^{\pm}$ can be taken into account. However, for completeness we have added also as hypothetical initial states the Bell states $\Phi^{\pm}$. We see that, just hypothetically, they would preserve concurrence $1$, although the process would be ``transparent'' for the state $\Phi^{+}$ in the sense that the first tree-level correction would result to be zero.}.

We consider now the last two processes involving photons. 
For the process of annihilation in two photons
we see that maximum concurrence is conserved for all the
values of the parameters only for
three Bell states as initial states, while $\Phi^{-}$
goes in a combination with maximal concurrence achieved only for some values of the parameters. 
For the Compton process we see that conservation of maximal entanglement is not allowed at all.
It is important to underline that in the case of the Compton scattering the interaction never generates a maximally entangled state as also found in Ref.\cite{CerveraLierta:2019ejf}.



\subsection{Entanglophilus, Entanglophobus and mixed regimes}
\label{ER}

Coming back to the two configurations $\mathcal{A}$ and $\mathcal{B}$ introduced in sec.\ref{sec3c}, we can identify the entanglophilus, entanglophobous and mixed regimes in a more systematic way. In order to characterize such patterns let us again restrict to the Bhabha scattering and, starting from the two pairs of Bell states in Eq.\eqref{BellStates}, we consider the states parameterized by the angles 
$\alpha$ and $\beta$ (see Eq.\eqref{instategen}) and defined by:
\be
\Phi^{\pm}_\alpha \equiv \cos{\alpha} \ket{RR} \pm \sin{\alpha} \ket{LL}\,; \qquad
\Psi^{\pm}_\beta \equiv \cos{\beta} \ket{RL} \pm \sin{\beta} \ket{LR}.
\ee
If, for low initial momenta, one lowers the $\alpha$-parameter from $\pi/4$ up to zero in $\Phi^{+}_\alpha$,   
it results that after the scattering the entanglement will be greater than its initial value for every $\theta$: this identifies the regime that we termed as \textit{entanglophilus}. Such behavior qualitatively persists 
also in the relativistic regime, although much attenuated. By performing a similar analysis for the state $\Phi^{-}_\alpha$, we find lower entanglement in the final state previous defined as \textit{entanglophobus} regime. 

\begin{figure}[t]
 \includegraphics[width =10 cm]{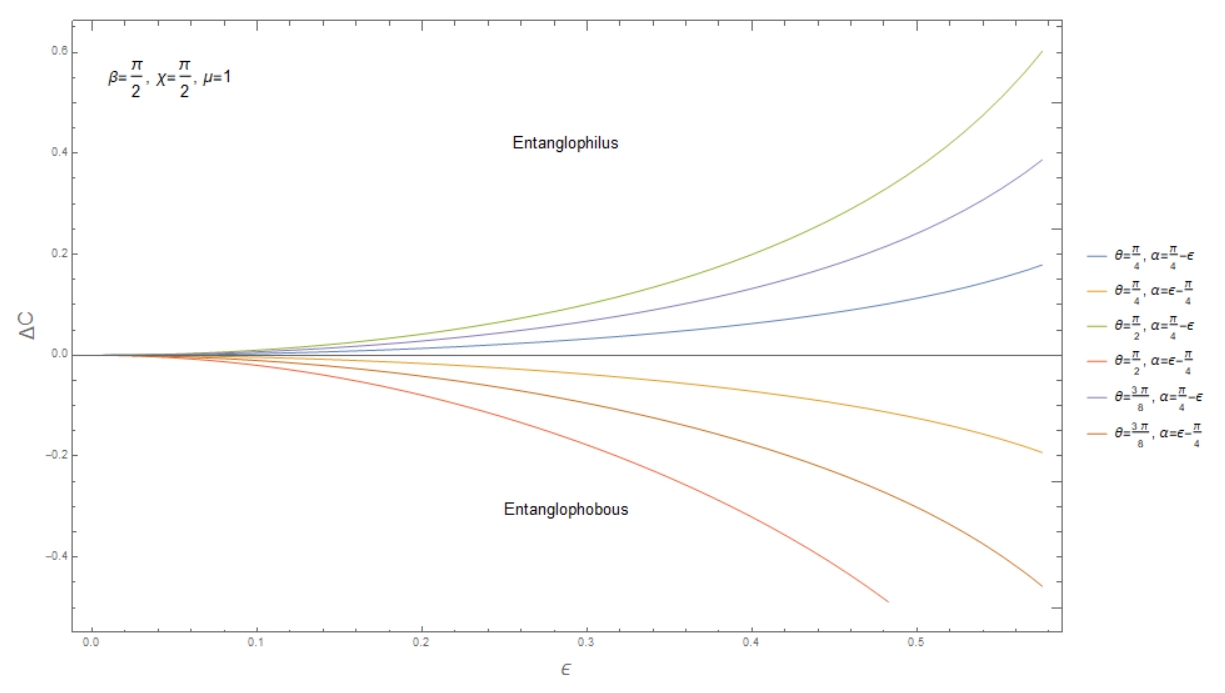}
 \caption{Plot of $\Delta{C}$ for  states $\Phi^{+}_\alpha$ (entanglophilus) and $\Phi^{-}_\alpha$ 
(entanglophobus) as a function of $\alpha$.}
 \label{figseven}
\end{figure}

The situation is summarized in Fig.\ref{figseven}, where we plot the relative difference 
between final and initial entanglement of states 
$\Phi^{+}_\alpha$ and $\Phi^{-}_\alpha$ defined as $\Delta{C}\equiv (C_f - C_i)/C_i$.

\begin{figure}[t]
 \subfloat[][\emph{}]{\includegraphics[width =6.55 cm]{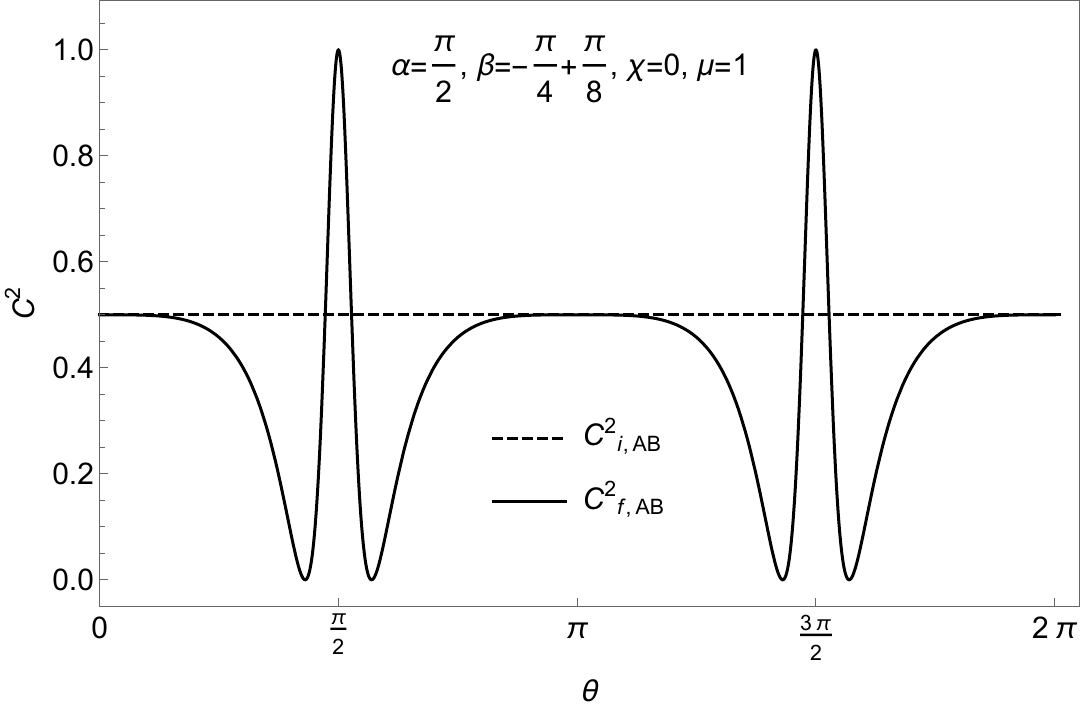}
\label{fig6a}}\quad
 \subfloat[][\emph{}]{\includegraphics[width =6.55 cm]{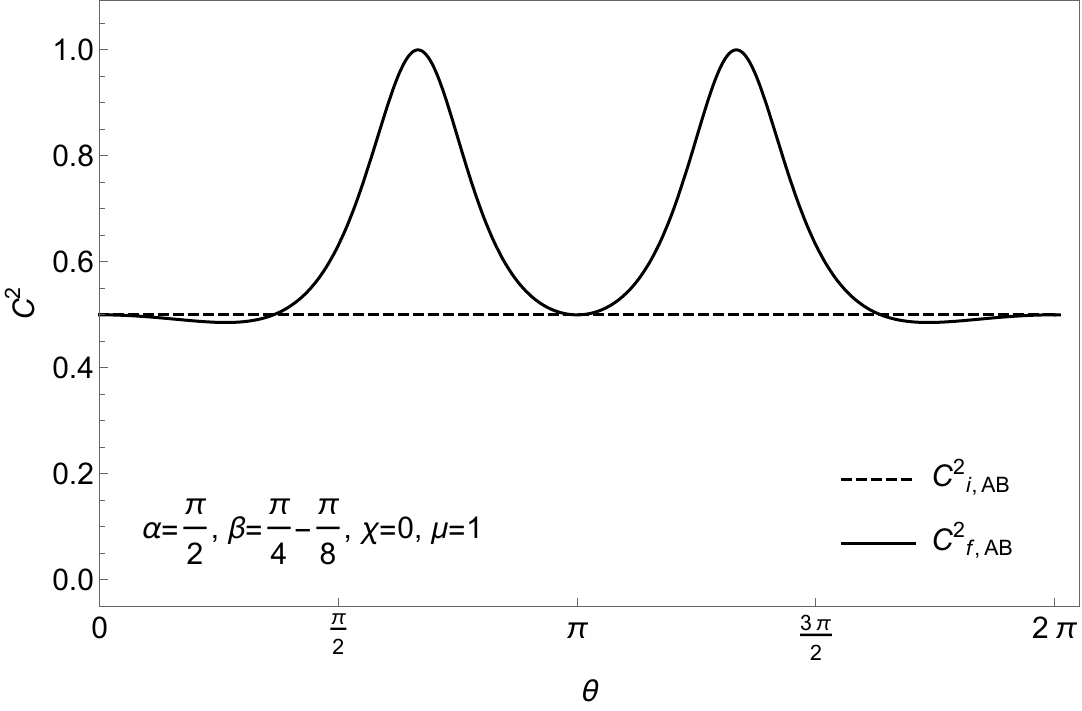}
\label{fig6b}}\quad
\caption{Plot of $\Delta{C}$ for  states $\Psi^{+}_\beta$  and $\Psi^{-}_\beta$  as a function of $\theta$. } 
\label{figsix}
\end{figure}

A similar analysis can be extended to the states $\Psi^{\pm}_\beta$. 
In this case we find an intermediate situation, the \textit{mixed} regime, between the last two. 
It is interesting to note that these cases, described in Fig.\ref{figsix}, 
are not symmetric under the change of sign in the $\beta$-parameter.

\section{Discussion}
\label{diss}

In light of these results, two questions naturally arise: Can we consider and quantify the interaction due to the scattering process as a valuable resource for generating and transferring entanglement? Secondly, how can these results be interpreted? In the following two subsections, we address these questions.

\subsection{Entanglement resource}
\label{entres}

In order to understand if the scattering process can be considered a good resource to create pairs of entangled particles we report, as a paradigmatic example, the analysis of the entanglement generation starting from the simplest cases $I$ and $I\!I$ at various energy regimes. To this aim,  we have to know how many particles will be scattered in that domains of scattering angle $\theta$ for which the entanglement is significantly different from zero. We can identify such domains from the plot previously reported. For what concerns the number of particles in output, we know that it is proportional to the differential cross section. As shown in Ref.\cite{Peskin:1995ev}, for a $2\rightarrow 2$ scattering process where all four masses involved are identical, in the COM reference frame the differential cross section results:
\begin{equation}
    \frac{d\sigma}{d\Omega} = \frac{|\mathcal{M}|^2}{64\pi^2E^2_{CM}}.
\end{equation}
So, in order to estimate and characterize the entanglement as a real resource, we have choosen to weight the three CCR terms with the number of particles in the $\theta$-domain of interest defining the weighted averages as:
\begin{equation}
\overline{Q^2} = \frac{1}{\bf{N}}\sum\limits_{rs}\int_{\mathcal{D}} |M(a,b;r,s)|^2 \, 
Q^2(\te) d\theta,
\label{cmedio}
\end{equation}
 where  $\bf{N}=\sum\limits_{rs}\int_{\mathcal{D}} |M(a,b;r,s)|^2d\theta$ and $Q = P,V,C$. 

\begin{figure}[t]
 \subfloat[][\emph{}]{\includegraphics[width =5.55 cm]{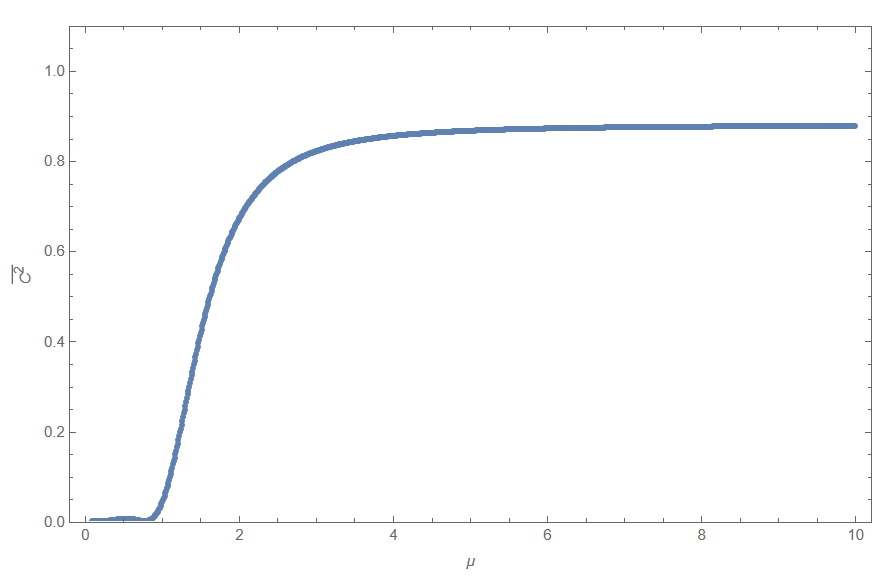}
\label{fig8a}}\quad
 \subfloat[][\emph{}]{\includegraphics[width =5.55 cm]{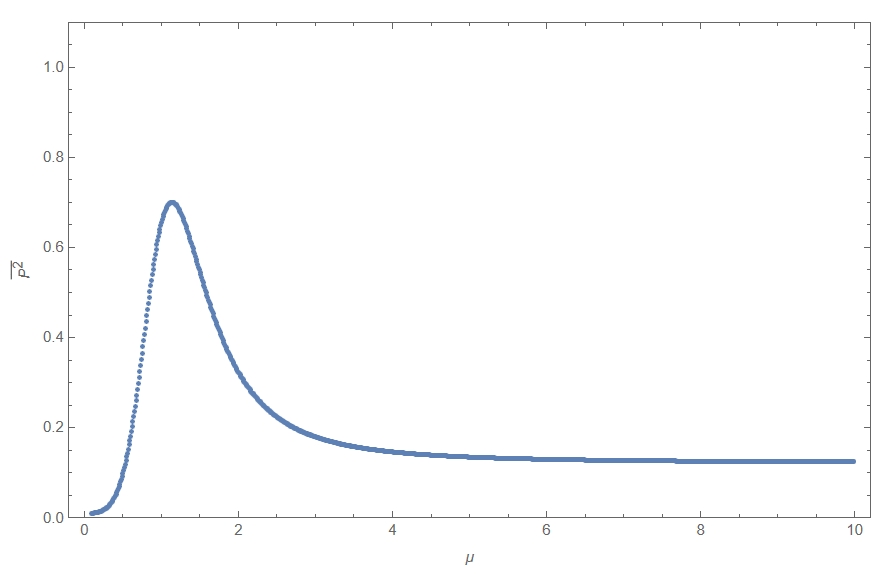}
\label{fig8b}}\quad
 \subfloat[][\emph{}]{\includegraphics[width =5.55 cm]{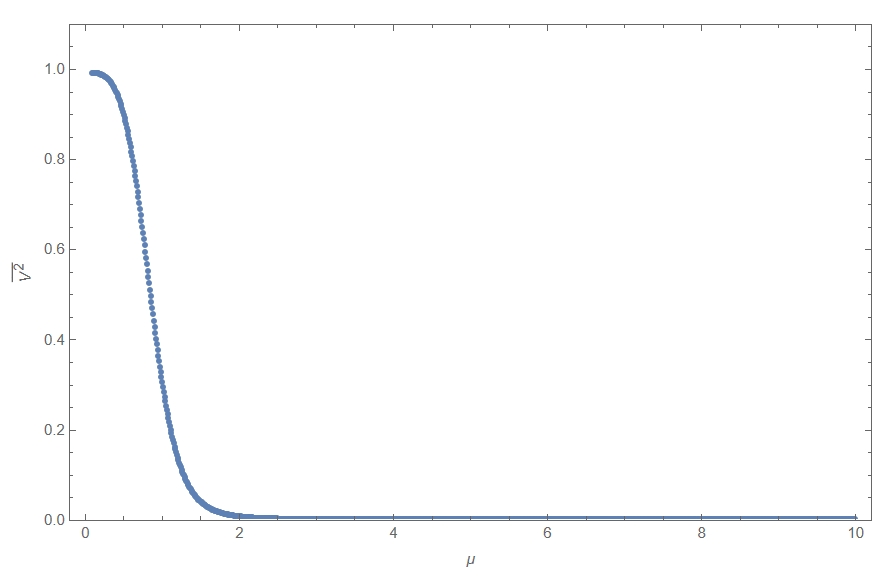}
\label{fig8c}}
\caption{Plot of the weighted average CCR terms as in definition Eq.\eqref{cmedio} in the domain $\mathcal{D}$, as functions of the incoming momentum $\mu$.} 
\label{figeight}
\end{figure}

Figs.\ref{fig8a}-\ref{fig8c} are relative to the initial state $\ket{RL}$. In this case, we have considered the relativistic limit for which, as shown in the plots \ref{fig1d}-\ref{fig1f}, the entanglement is significantly different from zero in a very sharp region around $\theta=\pi/2$. So, we averaged the CCR terms in a neighborhood of $\theta=\pi/2$ equal to $\mathcal{D}\equiv\{\pi/2-\pi/20, \pi/2+\pi/20\}$. It can be seen that if we narrow or widen the step from $\theta=\pi/2$ the concurrence  increases or decreases respectively, while predictability follows the reverse behavior. The triality relation Eq.\eqref{2} results verified also in this case.

\subsection{Interpretation}
\label{Disc}
Here we give a qualitative glimpse about possible symmetries 
hidden in the conservation of maximum entanglement, and provide
some analysis of the dynamics of the three quantum elements 
in the CCR in terms of probabilities and probability amplitudes.

We have seen that for scattering processes involving only
spin-$1/2$ massive particles, maximum concurrence is completely conserved,
while this is no more true for processes which include also
spin-$1$ massless particles. 
Let us then assume the following, somewhat suggestive, point of view.
The concurrence (see Eq. \ref{6}) can be interpreted as
the modulus of the determinant of the $2 \times 2$ matrix
\begin{equation}
M = \sqrt{2} \left(
\begin{array}{cc}
 a &  b \\
 c &  d
\end{array}
\right), 
\label{ConcDet}
\end{equation}
where $a, b, c, d$ are the coefficients
of the general two-qubit state Eq.\eqref{3}.
For a maximally entangled state, the modulus of the determinant is $1$ and 
it is simple to verify that the matrix $M$ becomes an orthogonal matrix.
The transformations $T$ that send an orthogonal matrix in another orthogonal matrix are the isometries of the Orthogonal Group.
For the first set of scattering processes, the output scattering amplitudes actually
send the coefficient's matrix of an initial maximally entangled state 
into the final state by an orthogonal transformation matrix \footnote{For example, in Bhabha 
scattering and for the Bell state $\Phi^{-}$,  
the final superposition results by an application on the initial matrix of coefficients of an orthogonal matrix $T$
with determinant $- 1$ (and argument $- s_1$ in the trigonometric
terms of the matrix),
while in M\o{}ller scattering and for the Bell state $\Phi^{+}$, the final superposition  
is obtained by applying an orthogonal matrix $T$
with determinant $1$ (in particular, a rotation of the angle $s_3$)}.
For processes involving photons, instead, this simple mechanism is partially,
or totally, lost. 

Moving to the analysis of the dynamics of CCR elements, we can write the latter in terms of probabilities and scattering amplitudes as 
\begin{eqnarray}
\label{ccrprob1}
    2P_{hs}(\rho_A)&=&P^2_A= (P_{RR} - P_{LL})^2 + (P_{RL} - P_{LR})^2 - 2 (P_{RR} - P_{LL}) (P_{LR} - P_{RL}).
    \\ 
\label{ccrprob2}
    2C_{hs}(\rho_A)&=&V^2_A = 4\biggl(P_{RR} P_{LR} + P_{LL} P_{RL} + 2 \sqrt{P_{RR} P_{LL} P_{RL} P_{LR}} \cos{(\xi + \eta - \tau)}\biggr)
\\
\label{ccrprob3}
     2C^{nl}_{hs}(\rho_{A|B})&=&C^2= 4\biggl(P_{RR} P_{LL} + P_{RL} P_{LR} - 2 \sqrt{P_{RR} P_{LL} P_{RL} P_{LR}} \cos{(\xi + \eta - \tau)}\biggr)
\end{eqnarray}
with similar expressions for  $P_{hs}(\rho_B)$ and $C_{hs}(\rho_B)$. Here the probability $P_{rs}, \; r, s = R, L$ is
obviously provided by the square modulus of the coefficient associated to the element $\ket{rs}$, while the phases
$\xi, \eta, \tau$ are associated to the terms $\ket{RL}, \ket{LR}, \ket{LL}$, respectively.

First of all, note that the predictability does not contain probability amplitudes but only
probabilities, as is reasonable being this term associated to absence of quantum interference. 
On the contrary, the two local and nonlocal coherence measures, visibility and concurrence,
contain amplitudes and phases just associated to quantum coherent interference, 
and the corresponding terms in the two quantities compensate each other
in the sum. 
A very interesting situation happens when $P_{RR} = P_{LL}$ and $P_{RL} = P_{LR}$: 
the predictability contribution Eq.\eqref{ccrprob1} disappears, 
and only visibility and entanglement are present. From Eqs.\eqref{ccrprob2} and \eqref{ccrprob3} we see that, when the cosine is equal to $1$, only the visibility contributions survive for any scattering angle. On the other hand, when the cosine is equal to $- 1$, the state is a maximally entangled state for any scattering angle. 

\begin{figure}[t]
 \subfloat[][\emph{}]{\includegraphics[width =5.55 cm]
{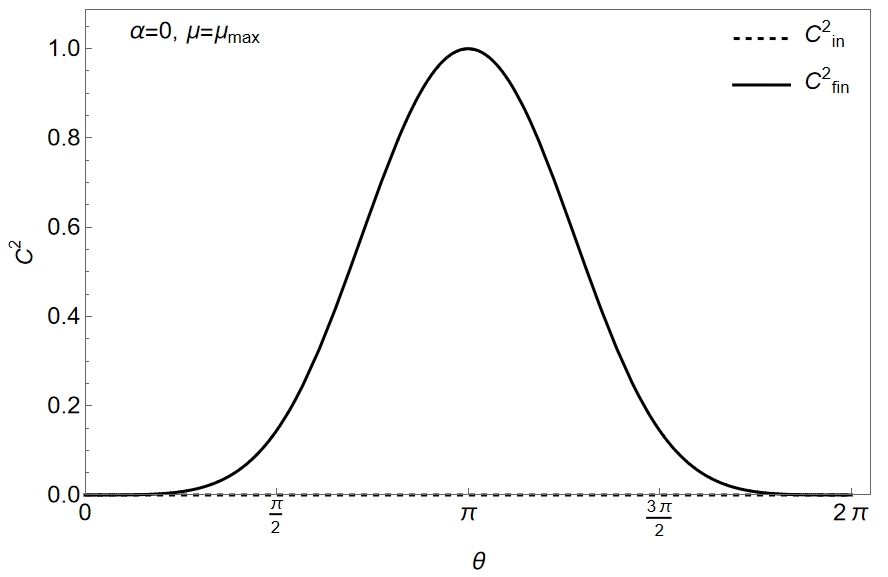}
\label{fig9a}}\quad
 \subfloat[][\emph{}]{\includegraphics[width =5.55 cm]{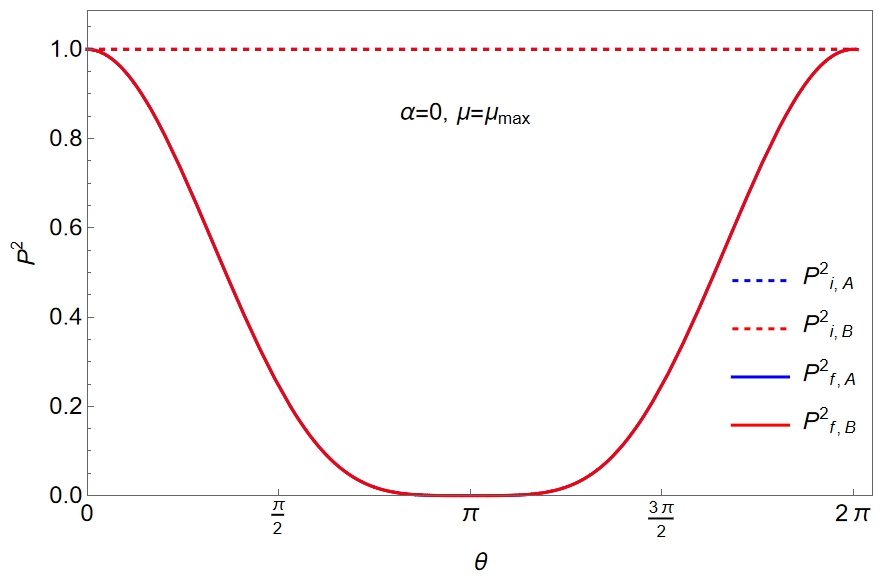}
\label{fig9b}}\quad
 \subfloat[][\emph{}]{\includegraphics[width =5.55 cm]{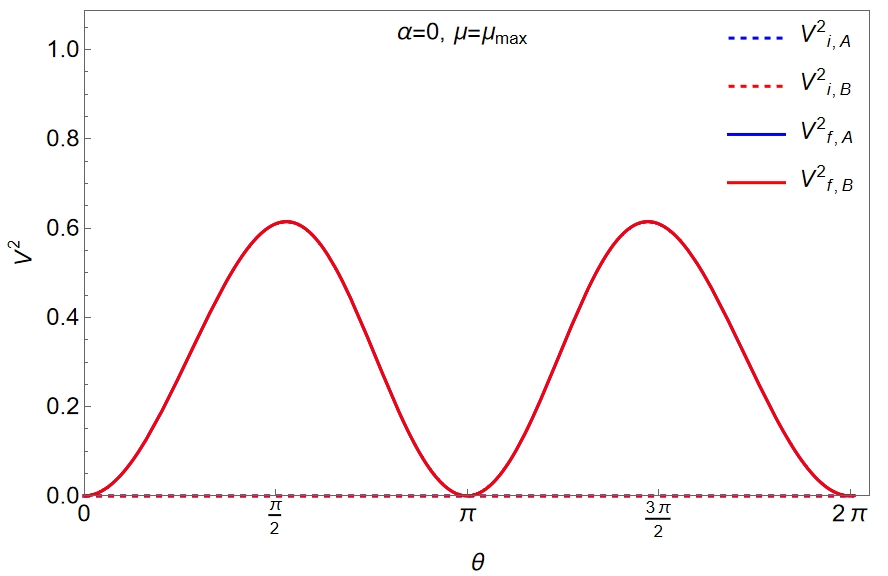}
\label{fig9c}}\quad
 \subfloat[][\emph{}]{\includegraphics[width =5.55 cm]{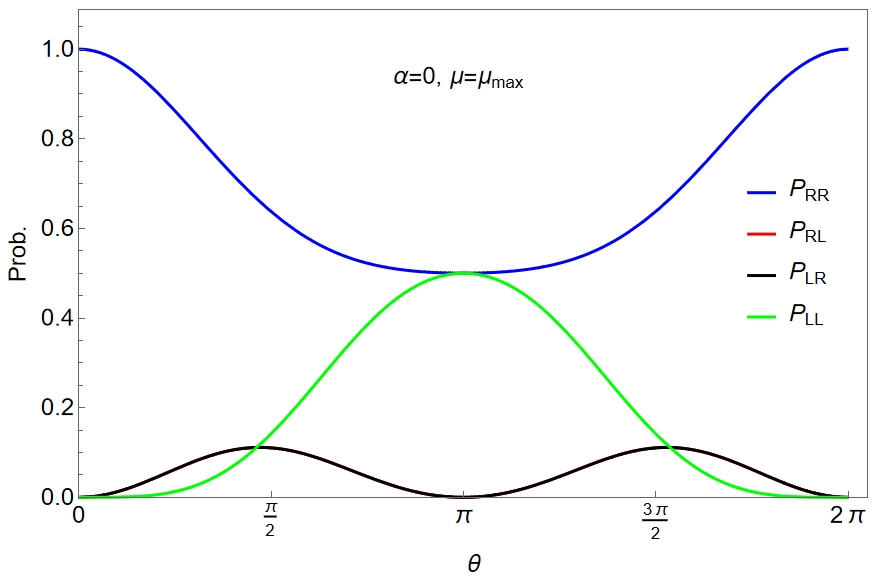}
\label{fig9d}}\quad
\caption{Plot of CCR terms with respect to the probabilities behaviour, after scattering for the initial state $\ket{RR}$ and $\mu=\mu_{max}$} 
\label{fignine}
\end{figure}

\begin{figure}[t]
 \subfloat[][\emph{}]{\includegraphics[width =5.55 cm]
{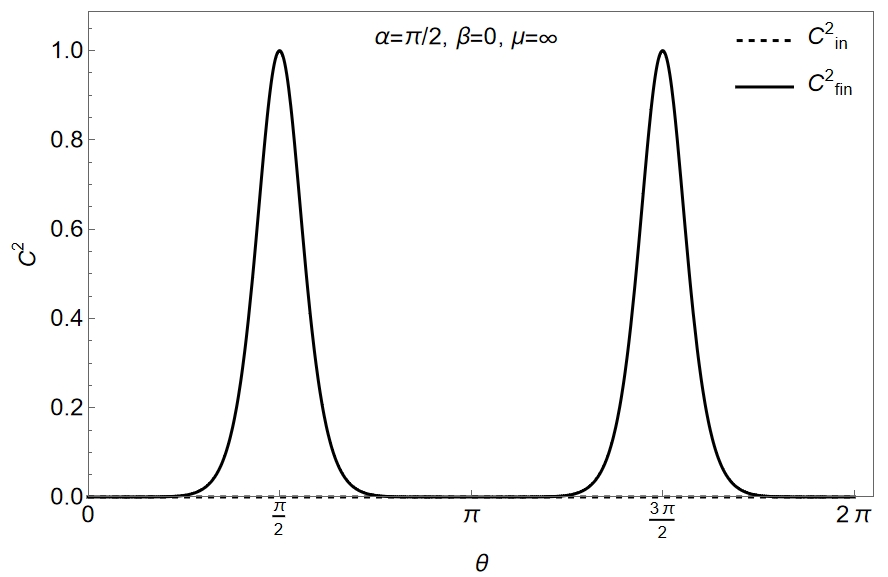}
\label{fig10a}}\quad
 \subfloat[][\emph{}]{\includegraphics[width =5.55 cm]{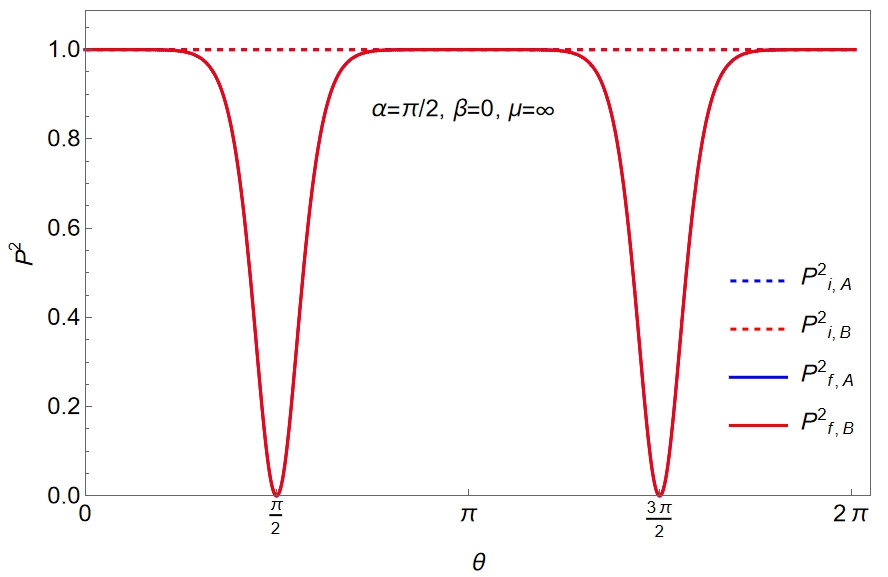}
\label{fig10b}}\quad
 \subfloat[][\emph{}]{\includegraphics[width =5.55 cm]{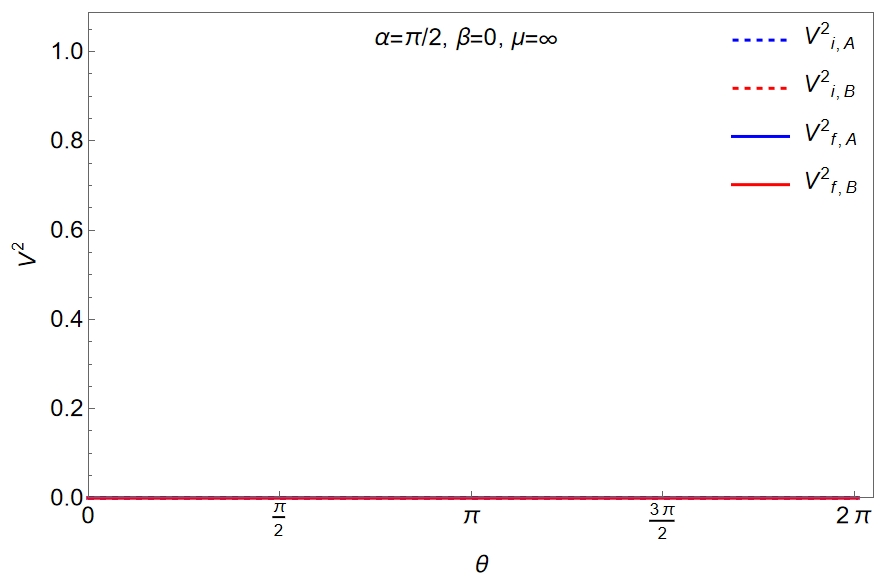}
\label{fig10c}}\quad
 \subfloat[][\emph{}]{\includegraphics[width =5.55 cm]
{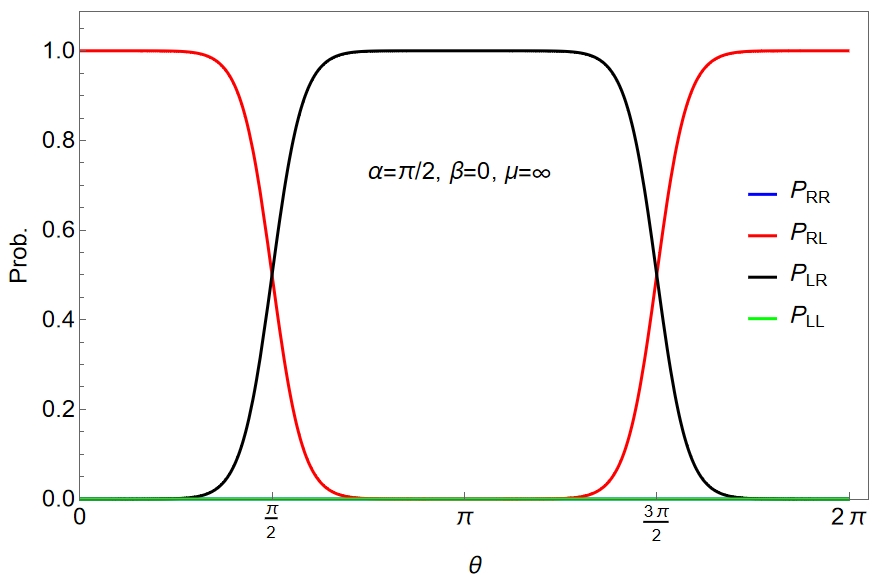}
\label{fig10c}}\quad
\caption{Plot of CCR terms with respect to the probabilities behaviour, after scattering for the initial state $\ket{RL}$ and $\mu=\infty$} 
\label{figten}
\end{figure}

This mechanism is also present for other values of the cosine, 
but only for particular values of the scattering angle. 
In fact, consider the simplest case in which the expressions of visibility and 
concurrence are given by
\begin{eqnarray}
    2C_{hs}(\rho_A)&=&V^2_A=4\biggl(P_{RR}P_{LR}+P_{LL}P_{RL}+2\mathcal{M}(RR)\mathcal{M}(RL)\mathcal{M}(LR)\mathcal{M}(LL)\biggr)
    \\
     2C^{nl}_{hs}(\rho_{A|B})&=&C^2=4\biggl(P_{RR}P_{LL}+P_{RL}P_{LR}-2\mathcal{M}(RR)\mathcal{M}(RL)\mathcal{M}(LR)\mathcal{M}(LL)\biggr).
\end{eqnarray}
In Fig.\ref{fignine} we plot the case relative to the initial state $\ket{RR}$ with $\mu = \mu_{max} \equiv \frac{1}{2}\sqrt{-3+\sqrt{17}}$, while in Fig.\ref{figten} the one relative to $\ket{RL}$ in the limit for $\mu=\infty$. As we can see, when the probabilities are equal pairwise in the following way: $P_{RR}=P_{LL}$ and $P_{RL}=P_{LR}$, then we obtain maximal entanglement and consistently $P=V=0$. In the plots Fig.\ref{fignine} and Fig.\ref{figten} we have reported the simplest cases corresponding to $P_{RR}=P_{LL}=1/2$, $P_{RL}=P_{LR}=0$ in $\theta = \pi$ and $P_{RL}=P_{LR}=1/2$, $P_{RR}=P_{LL}=0$ in $\theta = \pi/2,3\pi/2$, \smallskip respectively.

The analysis can be extended to the initial state $\ket{i}_{I\!I}$ and $\ket{i}_{I\!I\!I}$ as in Eqs.\eqref{instateII} and \eqref{instategen}. 
It turns out that the behaviour of probabilities are highly non-trivial, 
especially in the non-relativistic regime. Here, we report a particular initial configuration of parameter, 
that identify a sub-case of case II described by the state:

\begin{equation}
\label{inrefstate}
    \ket{i}_{REF} = \ket{R}_A\otimes\frac{1}{\sqrt{2}}\Big(\ket{R}_B+\ket{L}_B\Big).
\end{equation}

\begin{figure}[t]
 \subfloat[][\emph{}]{\includegraphics[width =5.55 cm]{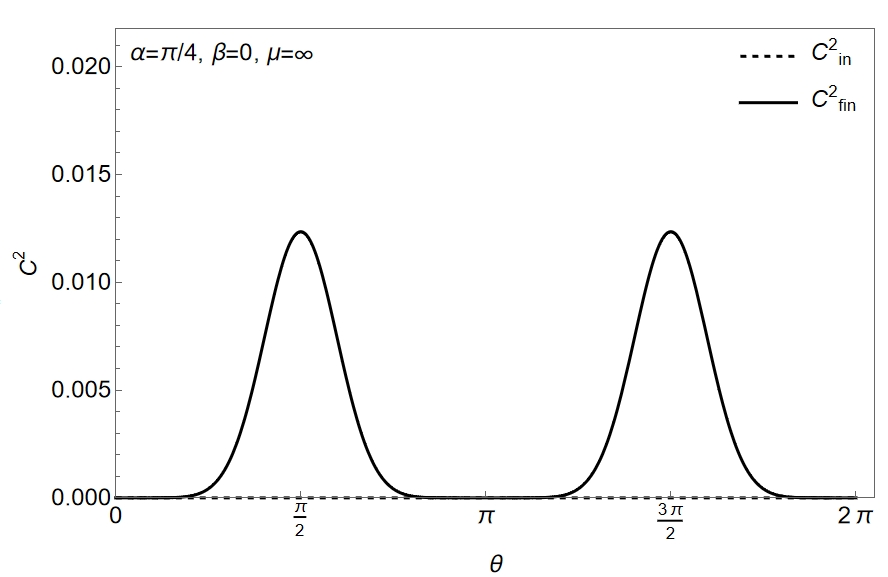}
\label{fig11b}}\quad
 \subfloat[][\emph{}]{\includegraphics[width =5.55 cm]{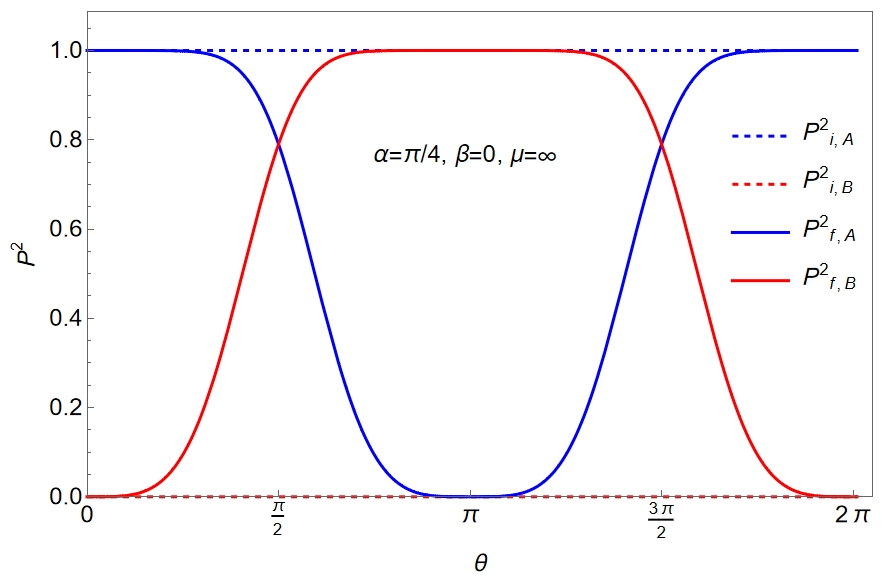}
\label{fig11c}}\quad
 \subfloat[][\emph{}]{\includegraphics[width =5.55 cm]{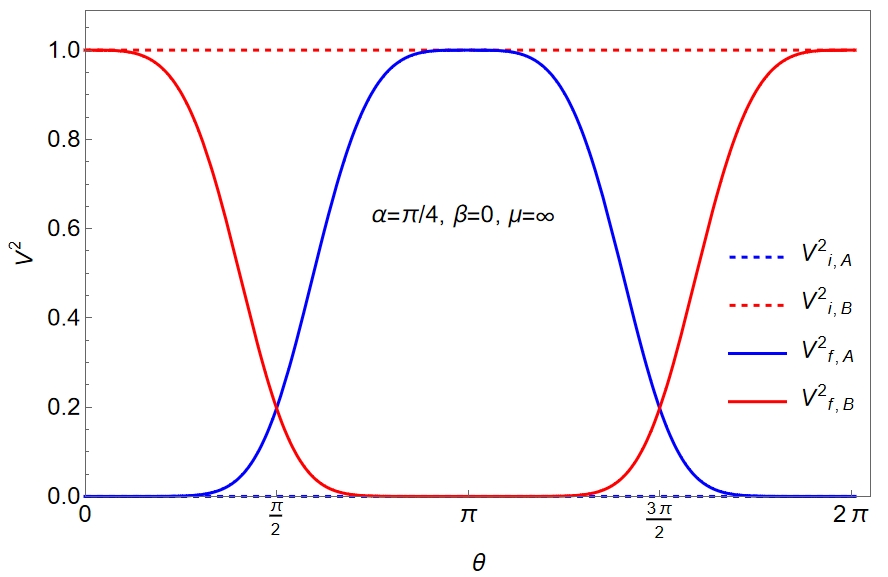}
\label{fig11c}}\quad
 \subfloat[][\emph{}]{\includegraphics[width =5.55 cm]{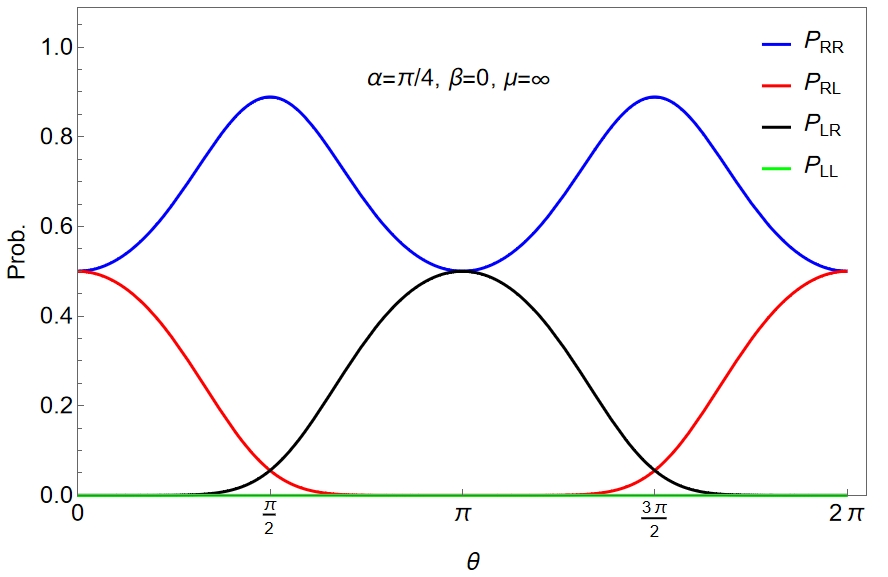}
\label{fig11a}}\quad
\caption{Plot of CCR terms with respect to the probabilities behaviour, after scattering for the initial state Eq.\eqref{inrefstate} and $\mu=\infty$.} 
\label{figeleven}
\end{figure}

As it can be seen from Fig. \ref{figeleven} and in analogy with the previous case, we see that the points in which visibility is maximal correspond to those for which probabilities 
are $P_{RR}=P_{LR}=1/2$ and $P_{RR}=P_{RL}=1/2$. Of course, when $P_{RL}=P_{LR}$ in $\theta = \pi/2,3\pi/2$  (Fig. \ref{fig11a}), 
even if the value of probabilities is low, the corresponding entanglement at such $\theta$ shows a peaks (Fig. \ref{fig11b}).

\section{Conclusions and outlook}
\label{conclusions}

In this work, we have analyzed the generation and transformation of entanglement 
in a QED scattering process at tree level, exploiting complete complementarity relations (CCR) 
which allow to take into account the intertwined evolutions of all quantum properties. 
Specifically, we focused on the Bhabha scattering process in the center-of-mass (COM) reference frame, 
where the particles are described by helicity states. We considered three distinct scenarios: 
the first, with the simplest factorized initial state; a second one, 
where the factorized initial state is expressed as a superposition of helicity states; 
and a third one, given by a more general case in which the particles are initially entangled in helicity.

The CCR, whose triality relation Eq.~\eqref{2} is verified for both the initial and final states, 
provide a comprehensive characterization of the quantum aspects of a system composed 
of interacting elementary particles. Within this framework, we examined the interplay 
among the three elements of CCR, finding that their evolution after scattering 
is multifaceted and non-trivial, depending on the choice of the initial parameters.

More precisely, for the simplest factorized initial state, where only predictability exists before scattering, 
the other two CCR terms emerge in the non-relativistic limit, 
while only entanglement develops in the relativistic regime. 
This behavior persists even if the initial state exhibits only visibility. 
In this scenario, the two partitions of the system---electron and positron---display 
identical CCR behavior, as shown by coinciding plots.

The situation looks different in the case of initial factorized states with local coherence contributions, 
due to presence of interference terms in the two bipartitions just from the beginning 
as described in Eq.~\eqref{instateII}. The CCR terms for the two bipartitions exhibit 
new non-trivial behaviors, resulting in asymmetry with respect to 
$\theta = \pi$ in the non-relativistic regime, although such a symmetry 
is again restored in the relativistic regime.

The most significant results arise in the context of the third case, 
where the incoming particles are entangled. Depending on the initial parameters, 
after scattering we can distinguish four different regimes. 
The first, which we define \textit{entanglophilus}, 
is characterized by increased final entanglement compared to the initial one 
for any scattering angle $\theta$. The second regime, which we term \textit{entanglophobous}, 
shows instead reduced final entanglement with respect to the initial one for any scattering angle $\theta$. 
The third, the \textit{mixed} regime, represents an intermediate situation between the last two, in which the increasing or decreasing of the final entanglement depend on the value of the scattering angle. 
These three patterns can be deduced starting 
from the two pairs of Bell states $\Phi^{\pm}$ and $\Psi^{\pm}$ Eq.~\eqref{BellStates}
and gradually reducing the initial entanglement. The relative phase between the two states $\Phi^{\pm}$ 
in the non-factorizable superposition determines which of the first two cases applies,
 while the same procedure for the pair $\Psi^{\pm}$ leads to mixed regime.

This led eventually to the fourth scenario, involving maximally entangled states,
 which has been investigated not only in reference to the Bhahba scattering process 
as in the previous cases, but also for other QED scattering processes. 
We found that, for processes involving only massive spin-$1/2$ particles, 
if the initial state is prepared as a Bell state, or more in general as a state with maximal concurrence, 
the particles will remain maximally entangled for any value of the scattering angle (and of incoming momenta), and regardless of how many scattering processes they undergo. 
The situation looks different if also massless, spin $1$ particles (photons) 
are present: depending on the process, only a partial conservation of maximal entanglement, 
or no conservation at all, is detected.
A concise characterization of these aspects for all the processes is provided in
Tables \ref{tab1}-\ref{tab3}, where the correspondences among initial and final states 
are reported for initial Bell states. All these behaviors are direct consequence 
of QED symmetries, a topic that we plan to explore further in a subsequent work \cite{WIP}.

To assess how the generation and transformation of entanglement could be exploited as a resource
to create entangled particle pairs, we have investigate the Bhabha scattering process 
as a paradigmatic example. We demonstrate that, for a suitable range of scattering angles 
of an angular aperture of $\pi/10$, if one starts from the factorized state $\ket{RL}$ and increases the incoming momentum, 
the square of the final concurrence stabilizes around an high value of about $0.8$. 
This is just one example showing that, by adjusting the angular resolution 
of the experimental apparatus, one can select particles with a relevant level of entanglement.

Finally, some interpretation of the behavior of the CCR terms evolved after the scattering in terms of 
probability and probability amplitudes has been provided. 

These results can have potential applications in experimental 
and technological contexts, and the study of CCR  
in the realm of fundamental interactions can serve as a probe 
for testing the properties of QFT theories and beyond.

As future developments, it will be very interesting to extend investigations 
of the above aspects to other interactions beyond QED, 
and to study their formulation in different Lorentz reference systems.

\section*{Appendix: QED scattering amplitudes}
\label{appA}

The scattering amplitudes are calculated in the COM reference frame of particles $A$ and $B$. In the following, $p_1=(\omega,0,0,|\vec{p}|)$ and $p_2=(\omega,0,0,-|\vec{p}|)$ are the incoming 4-momenta that lie along the $z$-axis, while $p_3=(\omega,|\vec{p}|\sin\theta,0,|\vec{p}|\cos\theta)$ and $p_4=(\omega,-|\vec{p}|\sin\theta,0,-|\vec{p}|\cos\theta))$ are the outgoing four-momenta lying along a direction that form an angle $\theta$ with respect to $z$-axis. $a,b,r,s$ are the spin indices, $\mu \equiv \frac{|\bf{p}|}{m_e}$, where $|\bf{p}|$ is the incoming momentum in the COM reference frame and $m_e$ the electron mass.
\vspace{0.5cm}
\begin{center}
    \textbf{Bhabha scattering}
\end{center}
\begin{equation}
\mathcal{M}=e^2\left[\bar{v}(b,p_2)\gamma^{\mu}u(a,p_1)\frac{1}{(p_1+p_2)^2}\bar{u}(r,p_3)\gamma_{\mu}v(s,p_4)-\bar{v}(b,p_2)\gamma^{\mu}v(s,p_4)\frac{1}{(p_1-p_3)^2}\bar{u}(r,p_3)\gamma_{\mu}u(a,p_1)\right]
\end{equation}
\bea
\mathcal{M}(RR;RR)&=&\mathcal{M}(LL;LL)=\frac{(2+11\mu^2+8\mu^4+2\cos{\theta}+\mu^2\cos{2\theta})\csc^2{(\frac{\theta}{2})}}{4\mu^2(1+\mu^2)}
\\ [1.5mm]
\mathcal{M}(RR;\prescript{RL}{LR}{})&=&-\mathcal{M}(LL;\prescript{RL}{LR}{})=-\frac{(1+\mu^2\cos\theta)\cot{(\frac{\theta}{2})}}{\mu^2\sqrt{1+\mu^2}}
\\ [1.5mm]
\mathcal{M}(RR;LL)&=&\mathcal{M}(LL;RR)=\frac{1+\mu^2(1+\cos{\theta})}{\mu^2(1+\mu^2)}
\\[1.5mm]
\mathcal{M}(\prescript{RL}{LR}{};RR)&=&-\mathcal{M}(\prescript{RL}{LR}{};LL)=\frac{(1+\mu^2\cos\theta)\cot{(\frac{\theta}{2})}}{\mu^2\sqrt{1+\mu^2}}
\\[1.5mm]
\mathcal{M}(RL;RL)&=&\mathcal{M}(LR;LR)=\frac{(1+\mu^2(1+\cos{\theta}))\cot^2{(\frac{\theta}{2})}}{\mu^2}
\\ [1.5mm]
\mathcal{M}(RL;LR)&=&\mathcal{M}(LR;RL)=1-\cos{\theta}-\frac{1}{\mu^2}
\eea
\vspace{0.5cm}
\begin{center}
    \textbf{M\o{}ller scattering}
\end{center}
\begin{equation}
\mathcal{M}=e^2\left[\bar{u}(r,p_3)\gamma^{\mu}u(a,p_1)\frac{1}{(p_1-p_3)^2}\bar{u}(s,p_4)\gamma_{\mu}u(b,p_2)-\bar{u}(r,p_3)\gamma^{\mu}u(b,p_2)\frac{1}{(p_2-p_3)^2}\bar{u}(s,p_4)\gamma_{\mu}u(a,p_1)\right]
\end{equation}
\bea
\mathcal{M}(RR;RR)&=&\mathcal{M}(LL;LL)=-\frac{(3+8\mu^2+\cos{2\theta})\csc^2{\theta}}{\mu^2}
\\ [1.5mm]
\mathcal{M}(RR;\prescript{RL}{LR}{})&=&\mathcal{M}(LL;\prescript{RL}{LR}{})=\mp\frac{2\sqrt{1+\mu^2}\cot{\theta}}{\mu^2}
\\ [1.5mm]
\mathcal{M}(RR;LL)&=&\mathcal{M}(LL;RR)=\frac{2}{\mu^2}
\\[1.5mm]
\mathcal{M}(RL;\prescript{RR}{LL}{})&=&-\mathcal{M}(LR;\prescript{RR}{LL}{})=\frac{2\sqrt{1+\mu^2}\cot{\theta}}{\mu^2}
\\[1.5mm]
\mathcal{M}(RL;RL)&=&\mathcal{M}(LR;LR)=-(2\cot^2\frac{\theta}{2}+\frac{1}{\mu^2}\cos\theta\csc^2\frac{\theta}{2})
\\ [1.5mm]
\mathcal{M}(RL;LR)&=&\mathcal{M}(LR;RL)=(2\tan^2\frac{\theta}{2}-\frac{1}{\mu^2}\cos\theta\sec^2\frac{\theta}{2})
\eea
\vspace{0.5cm}
\begin{center}
    \boldmath{$e^{-}e^{+}\rightarrow \mu^{-}\mu^{+}$}
\end{center}
\begin{equation}
\mathcal{M}=e^2\left[\bar{v}(b,p_2)\gamma^{\mu}u(a,p_1)\frac{1}{(p_1+p_2)^2}\bar{u}(r,p_3)\gamma_{\mu}v(s,p_4)\right]
\end{equation}
\bea
\mathcal{M}(RR;\prescript{RR}{LL}{})&=&\mathcal{M}(LL;\prescript{LL}{RR}{})=\mp\frac{\lambda\cos{\theta}}{\lambda^2+\mu^2}
\\ [1.5mm]
\mathcal{M}(RR;\prescript{RL}{LR}{})&=&-\mathcal{M}(LL;\prescript{RL}{LR}{})=\frac{\lambda\sin\theta}{\sqrt{\lambda^2+\mu^2}}
\\ [1.5mm]
\mathcal{M}(\prescript{RL}{LR}{};RR)&=&-\mathcal{M}(\prescript{RL}{LR}{};LL)=-\frac{\sin\theta}{\sqrt{\lambda^2+\mu^2}}
\\[1.5mm]
\mathcal{M}(RL;RL)&=&\mathcal{M}(LR;LR)=-(1+\cos\theta)
\\ [1.5mm]
\mathcal{M}(RL;LR)&=&\mathcal{M}(LR;RL)=(1-\cos\theta)
\eea
\vspace{0.5cm}
\begin{center}
    \boldmath{$e^{-}\mu^{-}\rightarrow e^{-}\mu^{-}$}
\end{center}
\begin{equation}
\mathcal{M}=e^2\left[\bar{u}(s,p_4)\gamma^{\mu}u(b,p_2)\frac{1}{(p_1-p_3)^2}\bar{u}(r,p_3)\gamma_{\mu}u(a,p_1)\right]
\end{equation}
\bea
\mathcal{M}(RR;RR)&=&\mathcal{M}(LL;LL)=-\frac{\mu^2(3-\cos\theta)+\sqrt{(1+\mu^2)(\lambda^2+\mu^2)}(1+\cos\theta)}{\mu^2(-1+\cos\theta)}
\\ [1.5mm]
\mathcal{M}(RR;RL)&=&-\mathcal{M}(LL;LR)=\frac{\sqrt{\lambda^2+\mu^2}\cot({\theta}/2)}{\mu^2}
\\ [1.5mm]
\mathcal{M}(RR;LR)&=&-\mathcal{M}(LL;RL)=-\frac{\lambda\sqrt{1+\mu^2}\cot({\theta}/2)}{\mu^2}
\\[1.5mm]
\mathcal{M}(RR;LL)&=&\mathcal{M}(LL;RR)=-\frac{\lambda}{\mu^2}
\\ [1.5mm]
\mathcal{M}(RL;RR)&=&-\mathcal{M}(LR;LL)=-\frac{\sqrt{\lambda^2+\mu^2}\cot({\theta}/2)}{\mu^2}
\\ [1.5mm]
\mathcal{M}(RL;LL)&=&-\mathcal{M}(LR;RR)=-\frac{\lambda\sqrt{1+\mu^2}\cot({\theta}/2)}{\mu^2}
\\[1.5mm]
\mathcal{M}(RL;RL)&=&\mathcal{M}(LR;LR)=\frac{\biggl(\mu^2+\sqrt{(1+\mu^2)(\lambda^2+\mu^2)}\biggr)\cot^2(\theta/2)}{\mu^2}
\\ [1.5mm]
\mathcal{M}(RL;LR)&=&\mathcal{M}(LR;RL)=\frac{\lambda}{\mu^2}
\eea
In the following we report the scattering amplitudes in the COM reference frame that involve photons as input and/or output state. In these cases the indices  $r$ and $s$ are associated to the polarization $\lambda=\pm 1$. We can define the photons states expressed in circular basis as: $\ket{R} \equiv \epsilon_{\lambda=+1}(\theta,\phi)=\frac{e^{i\phi}}{\sqrt{2}}\bigl(0,-\cos\theta\cos\phi+i\sin\phi,-\cos\theta\sin\phi-i\cos\phi,\sin\theta)$ and $\ket{L} \equiv \epsilon_{\lambda=-1}(\theta,\phi)=\frac{e^{-i\phi}}{\sqrt{2}}\bigl(0,\cos\theta\cos\phi+i\sin\phi,\cos\theta\sin\phi-i\cos\phi,-\sin\theta)$. In the scattering $e^{-}e^{+}\rightarrow \gamma\gamma$ we choose the outgoing photons momenta as $p_3 = (\omega,\omega\sin\theta,0,\omega\cos\theta)$ and $p_4 = (\omega,-\omega\sin\theta,0,-\omega\cos\theta)$. In the case of Compton process the momentum of the outgoing photon is $p_4 = (\omega,-\omega\sin\theta,0,-\omega\cos\theta)$.
\vspace{0.5cm}
\begin{center}
    \boldmath{$e^{-}e^{+}\rightarrow \gamma\gamma$}
\end{center}
\begin{equation}
\mathcal{M}=-e^2\epsilon_{\mu}^*(s,p_4)\epsilon_{\nu}^*(r,p_3)\bar{v}(b,p_2)\left[\frac{-\gamma^{\mu}p\!\!\!/_3\gamma^{\nu}+2\gamma^{\mu}p_1}{-2p_1\cdot p_3}+\frac{-\gamma^{\nu}p\!\!\!/_4\gamma^{\mu}+2\gamma^{\nu}p_1}{-2p_1\cdot p_4}  \right]u(a,p_1)
\end{equation}
\bea
\mathcal{M}(RR;RR)&=-&\mathcal{M}(LL;LL)=-\frac{4(\mu+\sqrt{1+\mu^2})}{\mu^2(1-\cos2\theta)+2}
\\ [1.5mm]
\mathcal{M}(RR;LL)&=-&\mathcal{M}(LL;RR)=\frac{4(-\mu+\sqrt{1+\mu^2})}{\mu^2(1-\cos2\theta)+2}
\\ [1.5mm]
\mathcal{M}(\prescript{RR}{LL}{};RL)&=&\mathcal{M}(\prescript{RR}{LL}{};LR)=\pm\frac{4\mu\sin^2\theta}{\mu^2(1-\cos2\theta)+2}
\\ [1.5mm]
\mathcal{M}(RL;\prescript{RR}{LL}{})&=&\mathcal{M}(LR;\prescript{RR}{LL}{})=0
\\ [1.5mm]
\mathcal{M}(RL;RL)&=&\mathcal{M}(LR;LR)=-\frac{2\mu\sqrt{1+\mu^2}(1+\cos\theta)\sin\theta}{1+\mu^2\sin^2\theta}
\\ [1.5mm]
\mathcal{M}(RL;LR)&=&\mathcal{M}(LR;RL)=\frac{2\mu\sqrt{1+\mu^2}(1-\cos\theta)\sin\theta}{1+\mu^2\sin^2\theta}
\eea
\vspace{0.5cm}
\begin{center}
    \textbf{Compton scattering}
\end{center}
\begin{equation}
\mathcal{M}=-e^2\epsilon_{\mu}^*(s,p_4)\epsilon_{\nu}(r,p_2)\bar{u}(b,p_3)\left[\frac{-\gamma^{\mu}p\!\!\!/_2\gamma^{\nu}+2\gamma^{\mu}p_1}{2p_1\cdot p_2}+\frac{-\gamma^{\nu}p\!\!\!/_4\gamma^{\mu}+2\gamma^{\nu}p_1}{-2p_1\cdot p_4}  \right]u(a,p_1)
\end{equation}
\bea
\mathcal{M}(RR;RR)&=&\mathcal{M}(LL;LL)=-\frac{4\mu\cos\theta/2+2(-\mu+\sqrt{1+\mu^2})\cos^3\theta/2}{\mu\cos\theta+\sqrt{1+\mu^2}}
\\ [1.5mm]
\mathcal{M}(RR;RL)&=&\mathcal{M}(LL;LR)=-\frac{(-\mu+\sqrt{1+\mu^2})\cos\theta/2}{\mu\cos\theta+\sqrt{1+\mu^2}}(1-\cos\theta)
\\ [1.5mm]
\mathcal{M}(RR;LR)&=&-\mathcal{M}(LL;RL)=\frac{1+\cos\theta}{\mu\cos\theta+\sqrt{1+\mu^2}}\sin\theta/2
\\ [1.5mm]
\mathcal{M}(RR;LL)&=&-\mathcal{M}(LL;RR)=\frac{2(-\mu+\sqrt{1+\mu^2})^2}{\mu\cos\theta+\sqrt{1+\mu^2}}\sin^3\theta/2
\\ [1.5mm]
\mathcal{M}(RL;RR)&=&\mathcal{M}(LR;LL)=-\frac{(-\mu+\sqrt{1+\mu^2})\cos\theta/2}{\mu\cos\theta+\sqrt{1+\mu^2}}(1-\cos\theta)
\\ [1.5mm]
\mathcal{M}(RL;LL)&=&-\mathcal{M}(LR;RR)=\frac{1+\cos\theta}{\mu\cos\theta+\sqrt{1+\mu^2}}\sin\theta/2
\\ [1.5mm]
\mathcal{M}(RL;RL)&=&\mathcal{M}(LR;LR)=-\frac{2(\mu+\sqrt{1+\mu^2})}{\mu\cos\theta+\sqrt{1+\mu^2}}\cos^3\theta/2
\\ [1.5mm]
\mathcal{M}(RL;LR)&=&-\mathcal{M}(LR;RL)=\frac{2\sin^3\theta/2}{\mu\cos\theta+\sqrt{1+\mu^2}}
\eea

\end{document}